\documentclass[12pt,pdflatex,sn-mathphys]{sn-jnl} 

\jyear{2022}%

\theoremstyle{thmstyleone}%
\theoremstyle{thmstyletwo}%

\theoremstyle{thmstylethree}%

\usepackage{siunitx} 
\usepackage{amsmath,amstext,color,tabu}
\usepackage{gensymb}
\usepackage{natbib}
\usepackage{hyperref}
\usepackage{longtable}
\usepackage{graphicx}
\usepackage{tablefootnote}
\def\ra#1#2#3{#1$^{\rm h}$#2$^{\rm m}$#3$^{\rm s}$}
\def\dec#1#2#3{$#1^\circ#2'#3''$}
\usepackage{setspace}
\begin{document}

\makeatletter
\let\jnl@style=\rm
\def\ref@jnl#1{{\jnl@style#1}}

\def\aj{\ref@jnl{Astron. J.}}                   
\def\actaa{\ref@jnl{Acta Astron.}}      
\def\araa{\ref@jnl{Annu. Rev. Astron. Astrophys.}}             
\def\apj{\ref@jnl{Astrophys. J.}}                 
\def\apjl{\ref@jnl{Astrophys. J.}}                
\def\apjs{\ref@jnl{ApJS}}               
\def\ao{\ref@jnl{Appl.~Opt.}}           
\def\apss{\ref@jnl{Ap\&SS}}             
\def\aap{\ref@jnl{Astron. Astrophys.}}                
\def\aapr{\ref@jnl{Astron. Astrophys.~Rev.}}          
\def\aaps{\ref@jnl{A\&AS}}              
\def\azh{\ref@jnl{AZh}}                 
\def\baas{\ref@jnl{BAAS}}               
\def\bac{\ref@jnl{Bull. astr. Inst. Czechosl.}}
\def\caa{\ref@jnl{Chinese Astron. Astrophys.}}
\def\cjaa{\ref@jnl{Chinese J. Astron. Astrophys.}}
\def\icarus{\ref@jnl{Icarus}}           
\def\jcap{\ref@jnl{J. Cosmology Astropart. Phys.}}
\def\jrasc{\ref@jnl{JRASC}}             
\def\memras{\ref@jnl{MmRAS}}            
\def\mnras{\ref@jnl{Mon. Not. R. Astron. Soc.}}             
\def\na{\ref@jnl{New A}}                
\def\nar{\ref@jnl{New A Rev.}}          
\def\pra{\ref@jnl{Phys.~Rev.~A}}        
\def\prb{\ref@jnl{Phys.~Rev.~B}}        
\def\prc{\ref@jnl{Phys.~Rev.~C}}        
\def\prd{\ref@jnl{Phys.~Rev.~D}}        
\def\pre{\ref@jnl{Phys.~Rev.~E}}        
\def\prl{\ref@jnl{Phys.~Rev.~Lett.}}    
\def\pasa{\ref@jnl{PASA}}               
\def\pasp{\ref@jnl{PASP}}               
\def\pasj{\ref@jnl{PASJ}}               
\def\rmxaa{\ref@jnl{Rev. Mexicana Astron. Astrofis.}}%
\def\qjras{\ref@jnl{QJRAS}}             
\def\skytel{\ref@jnl{S\&T}}             
\def\solphys{\ref@jnl{Sol.~Phys.}}      
\def\sovast{\ref@jnl{Soviet~Ast.}}      
\def\ssr{\ref@jnl{Space~Sci.~Rev.}}     
\def\zap{\ref@jnl{ZAp}}                 
\def\nat{\ref@jnl{Nature}}              
\def\iaucirc{\ref@jnl{IAU~Circ.}}       
\def\aplett{\ref@jnl{Astrophys.~Lett.}} 
\def\apspr{\ref@jnl{Astrophys.~Space~Phys.~Res.}}
\def\bain{\ref@jnl{Bull.~Astron.~Inst.~Netherlands}} 
\def\fcp{\ref@jnl{Fund.~Cosmic~Phys.}}  
\def\gca{\ref@jnl{Geochim.~Cosmochim.~Acta}}   
\def\grl{\ref@jnl{Geophys.~Res.~Lett.}} 
\def\jcp{\ref@jnl{J.~Chem.~Phys.}}      
\def\jgr{\ref@jnl{J.~Geophys.~Res.}}    
\def\jqsrt{\ref@jnl{J.~Quant.~Spec.~Radiat.~Transf.}}
\def\memsai{\ref@jnl{Mem.~Soc.~Astron.~Italiana}}
\def\nphysa{\ref@jnl{Nucl.~Phys.~A}}   
\def\physrep{\ref@jnl{Phys.~Rep.}}   
\def\physscr{\ref@jnl{Phys.~Scr}}   
\def\planss{\ref@jnl{Planet.~Space~Sci.}}   
\def\procspie{\ref@jnl{Proc.~SPIE}}   
\makeatother
\let\astap=\aap
\let\apjlett=\apjl
\let\apjsupp=\apjs
\let\applopt=\ao

\title[GRB\,211211A]{A Kilonova Following a Long-Duration Gamma-Ray Burst at 350~Mpc}


\author*[1]{\fnm{Jillian C.} \sur{Rastinejad}}\email{jillianrastinejad2024@u.northwestern.edu}
\author[2]{\fnm{Benjamin P.} \sur{Gompertz}}
\author[3]{\fnm{Andrew J.} \sur{Levan}}
\author[1]{\fnm{Wen-fai} \sur{Fong}}
\author[2]{\fnm{Matt} \sur{Nicholl}}
\author[4]{\fnm{Gavin P.} \sur{Lamb}}
\author[3,5,6]{\fnm{Daniele B.} \sur{Malesani}}
\author[1]{\fnm{Anya E.} \sur{Nugent}}
\author[2]{\fnm{Samantha R.} \sur{Oates}}
\author[4]{\fnm{Nial R.} \sur{Tanvir}}
\author[7]{\fnm{Antonio} \sur{de Ugarte Postigo}}
\author[1]{\fnm{Charles D.} \sur{Kilpatrick}}
\author[2]{\fnm{Christopher J.} \sur{Moore}}
\author[8,9]{\fnm{Brian D.} \sur{Metzger}}
\author[3,10]{\fnm{Maria Edvige} \sur{Ravasio}}
\author[11]{\fnm{Andrea} \sur{Rossi}}
\author[1]{\fnm{Genevieve} \sur{Schroeder}}
\author[12]{\fnm{Jacob} \sur{Jencson}}
\author[12]{\fnm{David J.} \sur{Sand}}
\author[12]{\fnm{Nathan} \sur{Smith}}
\author[13]{\fnm{Jos\'e Feliciano} \sur{Ag\"u\'i Fern\'andez }}
\author[14]{\fnm{Edo} \sur{Berger}}
\author[1]{\fnm{Peter K.} \sur{Blanchard}}
\author[15]{\fnm{Ryan} \sur{Chornock}}
\author[16]{\fnm{Bethany E.} \sur{Cobb}}
\author[17]{\fnm{Massimiliano} \sur{De Pasquale}}
\author[5,6]{\fnm{Johan P. U.} \sur{Fynbo}}
\author[18]{\fnm{Luca} \sur{Izzo}}
\author[13]{\fnm{D. Alexander} \sur{Kann}}
\author[3]{\fnm{Tanmoy} \sur{Laskar}}
\author[19]{\fnm{Ester} \sur{Marini}}
\author[1,20]{\fnm{Kerry} \sur{Paterson}}
\author[1]{\fnm{Alicia} \sur{Rouco Escorial}}
\author[1]{\fnm{Huei M.} \sur{Sears}}
\author[21]{\fnm{Christina C.} \sur{Th\"one}}

\affil*[1]{\orgdiv{Center for Interdisciplinary Exploration and Research in Astrophysics and Department of Physics and Astronomy}, \orgname{Northwestern University}, \orgaddress{\street{2145 Sheridan Road}, \city{Evanston}, \postcode{60208-3112}, \state{IL}, \country{USA}}}

\affil[2]{\orgdiv{Institute of Gravitational Wave Astronomy and School of Physics and Astronomy}, \orgname{University of Birmingham}, \orgaddress{\street{B15 2TT}, \country{UK}}}

\affil[3]{\orgdiv{Department of Astrophysics/IMAPP}, \orgname{Radboud University}, \orgaddress{\street{6525 AJ Nijmegen}, \country{The Netherlands}}}

\affil[4]{\orgdiv{School of Physics and Astronomy}, \orgname{University of Leicester}, \orgaddress{\street{LE1 7RH, University Road}, \city{Leicester}, \country{UK}}}

\affil[5]{\orgdiv{Cosmic Dawn Center (DAWN)}, \orgaddress{\country{Denmark}}}

\affil[6]{\orgdiv{Niels Bohr Institute}, \orgname{University of Copenhagen}, \orgaddress{\street{Jagtvej 128}, \postcode{2200}, \city{Copenhagen N}, \country{Denmark}}}

\affil[7]{\orgdiv{Artemis, Universit\'e C\^ote d'Azur, Observatoire de la C\^ote d'Azur}, \orgname{CNRS}, \orgaddress{\street{F-06304}, \city{Nice}, \country{France}}}

\affil[8]{\orgdiv{Center for Computational Astrophysics}, \orgname{Flatiron Institute}, \orgaddress{\street{162 W. 5th Avenue}, \city{New York}, \postcode{10011}, \state{NY}, \country{USA}}}

\affil[9]{\orgdiv{Department of Physics and Columbia Astrophysics Laboratory}, \orgname{Columbia University}, \orgaddress{\city{New York}, \postcode{10027}, \state{NY}, \country{USA}}}

\affil[10]{\orgdiv{INAF}, \orgname{Astronomical Observatory of Brera}, \orgaddress{\street{via E. Bianchi 46}, \postcode{23807}, \city{Merate (LC)}, \country{Italy}}}

\affil[11]{\orgdiv{INAF}, \orgname{Osservatorio di Astrofisica e Scienza dello Spazio}, \orgaddress{\street{via Piero Gobetti 93/3}, \postcode{40129}, \city{Bologna}, \country{Italy}}}

\affil[12]{\orgdiv{Steward Observatory}, \orgname{University of Arizona}, \orgaddress{\street{933 North Cherry Avenue}, \city{Tucson}, \postcode{85721-0065}, \state{AZ}, \country{USA}}}

\affil[13]{\orgname{Instituto de Astrof\'isica de Andaluc\'ia (IAA-CSIC)}, \orgaddress{\street{Glorieta de la Astronom\'ia s/n}, \postcode{18008}, \city{Granada}, \country{Spain}}}

\affil[14]{\orgdiv{Center for Astrophysics}, \orgname{Harvard \& Smithsonian}, \orgaddress{\street{60 Garden St.}, \city{Cambridge}, \state{MA}, \country{USA}}}

\affil[15]{\orgdiv{Department of Astronomy}, \orgname{University of California, Berkeley}, \orgaddress{\postcode{94720-3411}, \state{CA}, \country{USA}}}

\affil[16]{\orgdiv{Department of Physics}, \orgname{The George Washington University}, \orgaddress{\city{Washington}, \postcode{20052}, \state{DC}, \country{USA}}}

\affil[17]{\orgdiv{Department of Mathematical, Informatics, Physical and Earth Sciences}, \orgname{Polo Papardo, University of Messina}, \orgaddress{\street{via F.S. D'Alcontres 31}, \city{Messina}, \postcode{98166}, \country{Italy}}}

\affil[18]{\orgdiv{Dark Cosmology Centre}, \orgname{University of Copenhagen}, \orgaddress{\street{Jagtvej 128}, \postcode{2200}, \city{Copenhagen N}, \country{Denmark}}}

\affil[19]{\orgdiv{INAF}, \orgname{Observatory of Rome}, \orgaddress{\street{Via Frascati 33}, \postcode{00077}, \city{Monte Porzio Catone (RM)}, \country{Italy}}}

\affil[20]{\orgdiv{Max-Planck-Institut f\"ur Astronomie (MPIA)}, \orgaddress{\street{K\"onigstuhl 17}, \postcode{69117}, \city{Heidelberg}, \country{Germany}}}

\affil[21]{\orgname{Astronomical Institute of the Czech Academy of Sciences (ASU-CAS)}, \orgaddress{\street{Fri\v cova 298}, \city{Ond\v rejov}, \postcode{251 65}, \country{Czech Republic}}}

\abstract{\doublespacing Gamma-ray bursts (GRBs) are divided into two populations \cite{Norris+84,Kouveliotou+93}; long GRBs that derive from the core-collapse of massive stars \cite[e.g.,][]{Galama+98} and short GRBs that form in the merger of two compact objects \cite{Abbott+17c,Goldstein+2017}. While it is common to divide the two populations at a $\gamma$-ray duration of two seconds, classification based on duration does not always map to the progenitor. Notably, GRBs with short ($\lesssim 2$ seconds) spikes of prompt $\gamma$-ray emission followed by prolonged, spectrally-softer extended emission (EE-SGRBs) have been suggested to arise from compact object mergers \cite{Norris02,NorrisBonnell06,Gehrels+06}. Compact object mergers are of great astrophysical importance as the only confirmed site of rapid neutron capture ($r$-process) nucleosynthesis, observed in the form of so-called kilonovae \cite{Arcavi+17,Coulter+17,Lipunov+17,Tanvir+17,Soares-Santos+17,Valenti+17}. Here, we report the discovery of a likely kilonova associated with the nearby (350 Mpc), minute-duration GRB\,211211A. The kilonova implies that the progenitor is a compact object merger, suggesting that GRBs with long, complex light curves can be spawned from merger events.
GRB\,211211A's kilonova has a similar luminosity, duration and color to that which accompanied the gravitational wave (GW)-detected binary neutron star (BNS) merger GW170817 \cite{Abbott+17c}. Further searches for GW signals coincident with long GRBs are a promising route for future multi-messenger astronomy.
}

\maketitle

\begin{figure}
\centerline{
\includegraphics[width=\textwidth]{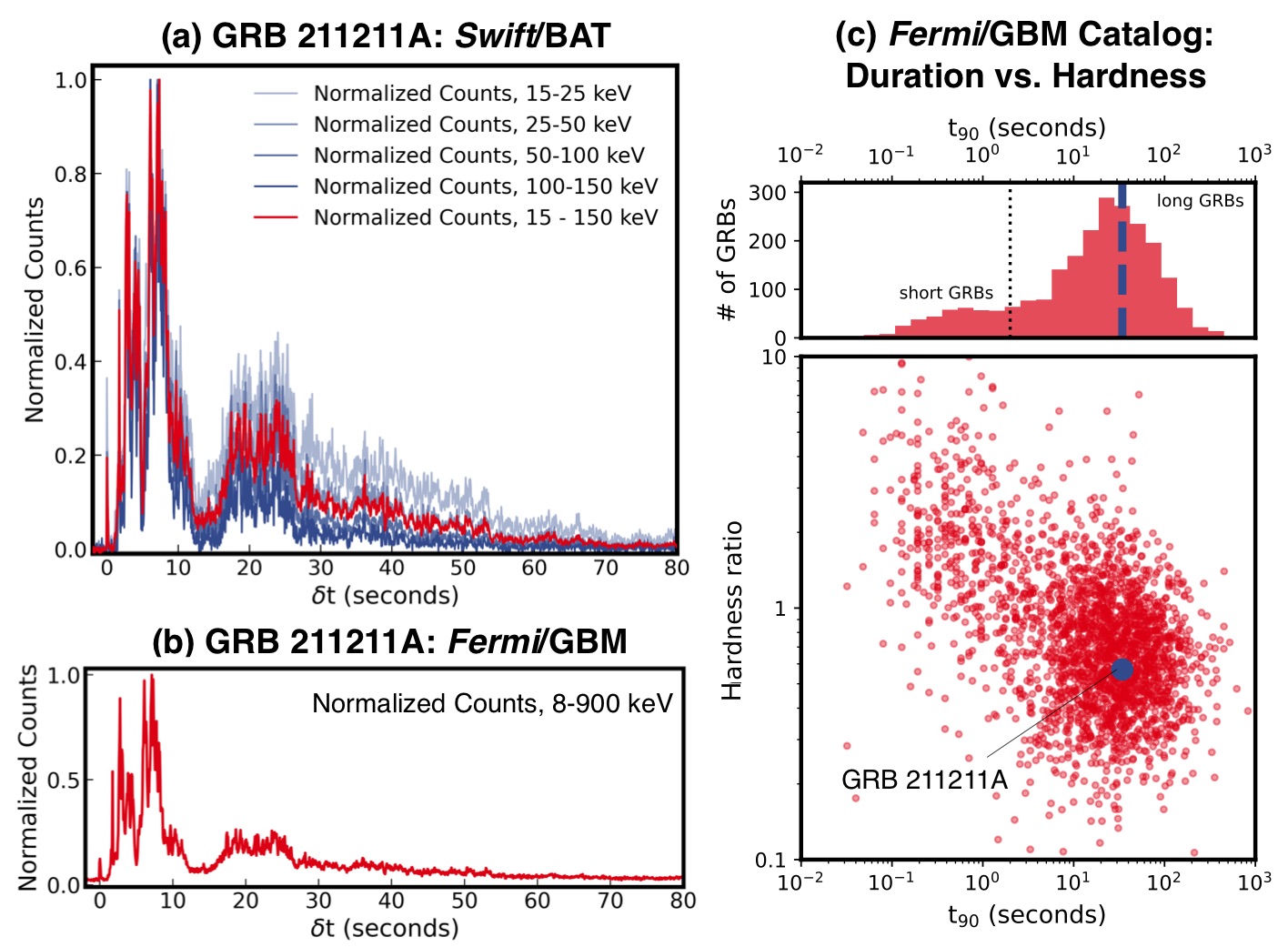}}
\caption{Figure 1.}
\label{fig:GRB_lc}
\end{figure}

On 2021 December 11 at 13:09 UT, the \textit{Neil Gehrels Swift Observatory's} ({\it Swift}) Burst Alert Telescope (BAT) identified the bright GRB\,211211A. The burst was discovered simultaneously by the \emph{Fermi} Gamma-ray Burst Monitor (GBM). The burst's duration of 51.37 $\pm$ 0.80~s \cite{GCN_211211A_BATT90} ($\sim 34.3$~s in GBM \cite{GCN_211211A_GBMT90}) and spectral hardness lie close to the mean of the long-GRB population (Figure~\ref{fig:GRB_lc}). The burst's light curve consists of several overlapping pulses exhibiting little spectral evolution and lasting for approximately 12~s, followed by longer-lived, and apparently softer emission extending to 50~s. Though GRB\,211211A's lack of early spectral evolution and later softening is reminiscent of the behaviour of past EE-SGRBs, these durations are far beyond those considered in previous searches for EE-SGRBs \cite{NorrisBonnell06,Kaneko+15}. The {\it Swift} X-ray Telescope (XRT) and Ultra-Violet Optical Telescope (UVOT) began observing the accompanying broadband afterglow $\sim 1$~minute after the burst (see Methods section `\textit{Swift} Observations').

\begin{figure*}
\centerline{
\includegraphics[width=\textwidth]{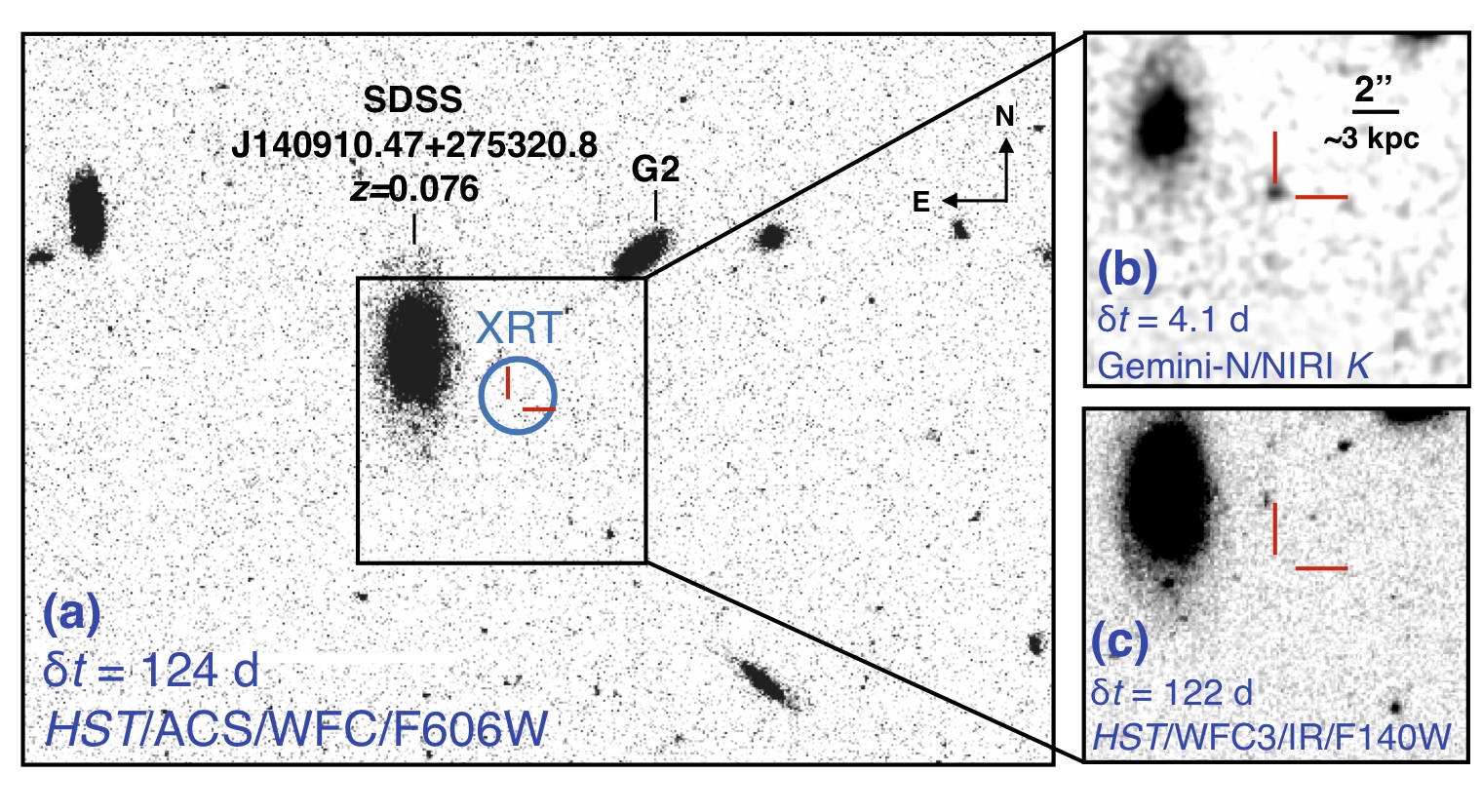}}
\caption{Figure 2.}
\label{fig:Hubble}
\end{figure*}

Motivated by GRB\,211211A's gamma-ray light curve and its proximity to the bright ($r = 19.4$~mag) galaxy SDSS J140910.47+275320.8 (Figure~\ref{fig:Hubble}), we initiated multi-wavelength follow-up observations. We obtained spectroscopy at the Nordic Optical Telescope (NOT; later confirmed with a Keck II spectrum, see Methods section `Host Galaxy Observations') that revealed the nearby galaxy is at a redshift $z=0.0763 \pm 0.0002$ (distance $\approx$350 Mpc). The modest offset between the galaxy and optical afterglow ($5.\!\!''44 \pm 0.\!\!''02$; 7.91 $\pm$ 0.03~kpc in projection), their low probability of chance coincidence (1.4\%, \cite{Bloom+02}), and the absence of any fainter, underlying host galaxy in late-time {\em Hubble Space Telescope} ({\em HST}) imaging provide compelling evidence that GRB\,211211A originated in SDSS J140910.47+275320.8 (Figure~\ref{fig:Hubble}). At 350~Mpc, GRB\,211211A is one of the closest bursts across both short and long classes discovered to date.

\begin{figure*}
\centerline{
\includegraphics[width=.8\textwidth]{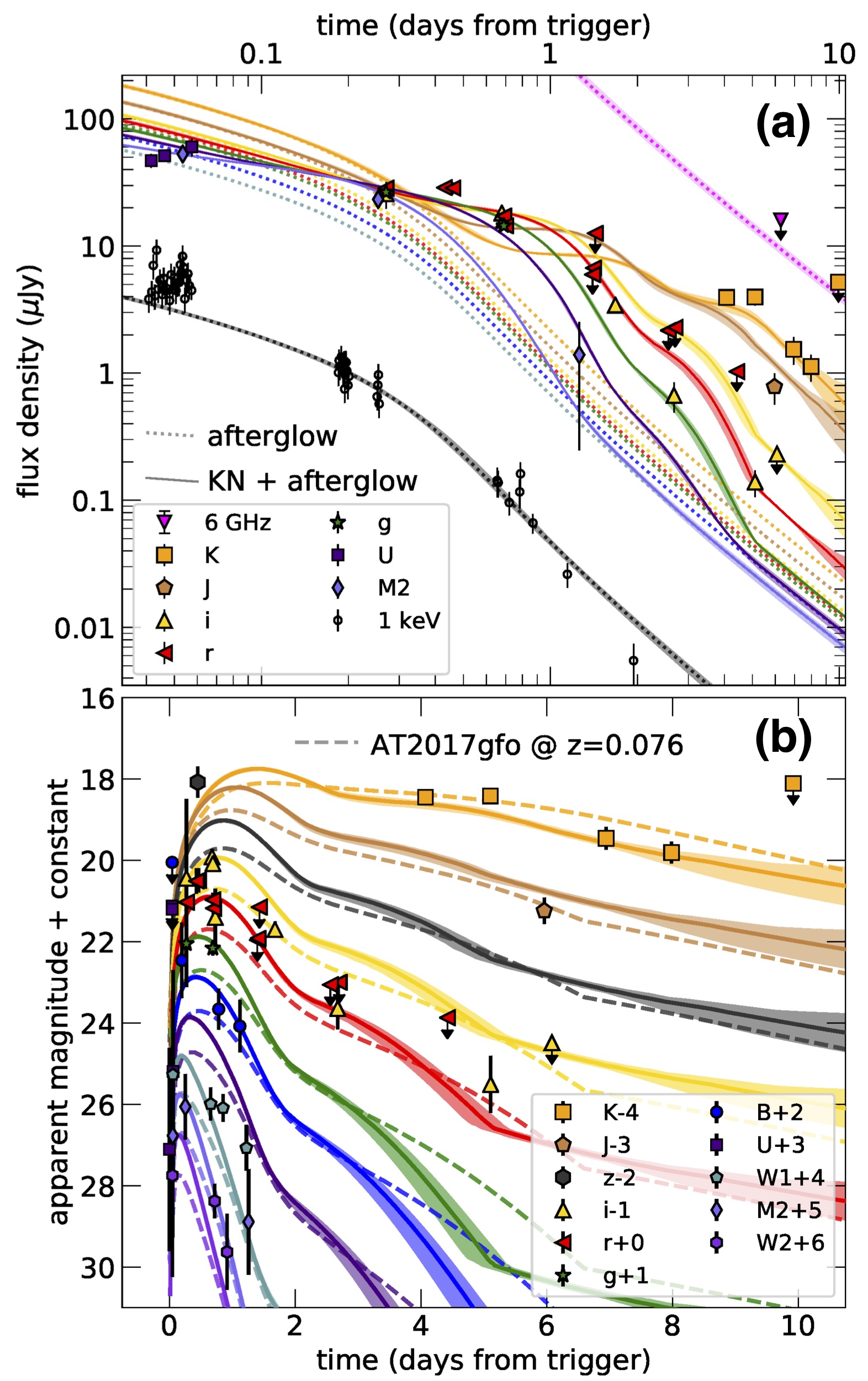}}
\caption{Figure 3.}
\label{fig:lc_models}
\end{figure*}

We obtained optical imaging with the NOT and the Calar Alto Observatory (CAHA) that showed an uncatalogued source fading rapidly over the first three days post-burst. At 4.1 days we observed in $K$-band with Gemini-North, detecting a $K = 22.4$~mag source, indicative of a strong IR excess compared to the optical afterglow light curve. We continued to observe in the $iJK$-bands with Gemini-North and the MMT to 10 days post-burst. At 6.3 days, we obtained a deep limit on the 6 GHz radio afterglow with the Karl Jansky Very Large Array (VLA). We acquired late-time optical and near-IR (NIR) observations with Gemini-North, MMT, the Large Binocular Telescope (LBT), Gran Telescopio Canarias (GTC) and {\em HST}. We obtained NOT imaging at 17.7 days post-burst that constrains an associated supernova (SN) to deep limits ($\nu L_{\nu} < 3 \times 10^{40}$~erg~s$^{-1}$, or $M_I$ $> -13$ mag). This rules out a typical long GRB massive star origin to limits a factor of $>200$ fainter than the proto-type GRB-SN 1998bw (assuming $z=0.076$; \cite{Galama+98}).
We present the full optical-NIR dataset for GRB\,211211A's counterpart in Extended Data Table~\ref{tab:observations} and describe the data reduction and analysis further in Methods (see sections `Optical Afterglow Observations' and `Further Optical-NIR Observations'). Notably, in $K$-band the luminosity at 4.1 days post-burst is approximately that of AT\,2017gfo ($\nu L_{\nu} \approx 8 \times 10^{40}$~erg s$^{-1}$) and the light curve fades at a remarkably similar rate to AT\,2017gfo (Figure~\ref{fig:lc_models}).

We first fit an afterglow model following the methods of \cite{Lamb+21} (and references therein; see Methods section `Afterglow Modeling') to the full X-ray and radio light curves, and to the UV-optical-NIR photometry at $\delta t < 0.1$~day (where $\delta t$ denotes the time after the BAT trigger), when the afterglow is expected to dominate any thermal counterpart. We find an isotropic-equivalent kinetic energy, $E_{\rm K,iso} = 5 \times 10^{52}$~erg. Other properties are listed in Extended Data Table~\ref{tab:ag_params}, and
are consistent with those inferred for other short GRBs. The $K$-band observation at 4.1 days is in excess of $\approx 3.8$~mag (a factor of 33 in brightness) compared to the corresponding model afterglow flux, which is well-constrained by the X-ray and radio data, necessitating an additional component in our model.

We thus obtain optical-NIR photometry after the subtraction of the afterglow component and considering the uncertainty in the afterglow model (Extended Data Table~\ref{tab:observations}). We fit the afterglow-subtracted photometry with a three-component kilonova model following \cite{villar+17,Nicholl+21} (see Methods section `Kilonova Modeling'). Our fitting indicates a total $r$-process ejecta mass of $M_{\rm ej} = 0.047^{+0.026}_{-0.011}\,{\rm M}_\odot$. This includes $\approx 0.02\,{\rm M}_\odot$ of lanthanide-rich (``red'') ejecta with velocity $v\approx 0.3c$ and $\approx 0.01\,{\rm M}_\odot$ of intermediate-opacity (``purple'') ejecta with $v\approx 0.1c$. Red ejecta can be produced in dynamical tides \cite{sekiguchi+16} or by winds from a remnant accretion disk if neutrino irradiation is low \cite{metzgerfernandez14}, though the high velocity found by our model is more consistent with a tidal origin. The purple ejecta are consistent with a disk wind, assuming moderate neutrino irradiation to lower the lanthanide fraction. The remaining $\approx 0.01\,{\rm M}_\odot$ is lanthanide-free (``blue'') material with $v\approx 0.3c$. This can be produced by dynamical shocks \cite{Bauswein2013}, winds from a long-lived magnetized NS (magnetar) remnant \cite{Metzger+18}, or from a disk wind with high neutrino irradiation. The blue ejecta mainly produce optical emission on timescales of $\sim1$ day, and hence are somewhat degenerate with early shock cooling of matter heated by the GRB jet \cite{Piro+18} (see Methods section `Kilonova Modeling'). Overall our best-fit masses are in reasonable agreement with estimates for AT\,2017gfo, though the reddest ejecta appear to be more massive in this case (Extended Data Figure 8). If we assume that the progenitor binary consists of two NSs and use predictions from merger simulations to constrain the relative component masses and velocities \cite{Nicholl+21}, we obtain a good fit with a 1.4+1.3 M$_{\odot}$ binary producing $\approx0.02$ M$_{\odot}$ of ejecta, though matching the luminosity in the first day may require additional heating by the GRB jet over the minute-long timescale of the burst (see Methods  section `Kilonova Modeling'; Extended Data Figures 4 and 5).

Though the profile of the initial gamma-ray pulse complex has a duration $\gg 2$~s, there are additional lines of evidence (beyond the kilonova), which link GRB\,211211A to a compact object merger. First, the observed exponential decline in X-rays at a few hundred seconds after trigger is a notable feature of EE-SGRBs \cite{Gompertz13}, and is highly consistent with both the luminosity and timescale of previous examples. Second, the spectral lag during the initial burst of $4 \pm 9$\,ms between the 25 -- 50 and 100 -- 150\,keV BAT bands is more consistent with short than long GRBs (see Methods section `Alternative Interpretations'; \cite{Bernardini15}). The host galaxy stellar population has mass $\approx 7\times10^8$~M$_\odot$ and star formation rate (SFR) $\approx 0.07$~M$_\odot$yr$^{-1}$ (specific SFR of $\approx 0.10$~Gyr$^{-1}$), which are more consistent with the hosts of short than long GRBs (see Methods section `Stellar Population Modeling') \cite{BRIGHT_II,Perley2013}. The host offset and the lack of any underlying stellar component can be readily explained in merger scenarios, but would be extremely unusual for a massive star. 

It is also relevant to consider if the counterpart could instead arise from a core-collapse or Ni-powered event, either at $z=0.076$, or from a more distant, and as yet unseen, galaxy. Importantly, our NOT limit of $i > 24.7$~mag rules out any known GRB supernova (SN) at $z=0.076$ and would have been sensitive to a GRB-SNe out to $z\approx0.5$ (see Methods section `Alternative Interpretations'). GRB\,211211A's $K$-band peak and temporal evolution are not compatible with any SN in the sample of \cite{Cano14}, including SN\,2010bh \cite{Olivares+12}, the dimmest, fastest-fading GRB-SNe considered. There is also no sign of significant stellar mass or star formation at the burst location that might obscure the SN with dust (Figure~\ref{fig:Hubble}), nor is there evidence of significant absorption in the host spectral energy distribution which extends to $<2000$~\AA. Additionally, light curve models powered by $^{56}$Ni decay (relevant for an SN, or possibly a merger between a NS and white dwarf \cite{King2007}) are unable to provide satisfactory fits to our data (Methods section '$^{56}$Ni-Powered Transient Model', Extended Data Figure~6). Higher-redshift scenarios are limited by multiple observational constraints. First, the detection in the {\em Swift}-$uvw2$ band demonstrates no absorption from neutral hydrogen at 1928~\AA{} (observed), implying $z<1.4$ (99\% confidence level). Furthermore, deep {\em HST} observations reach $\mathrm{F606W} > 27.8$~mag and $\mathrm{F140W} > 27.2$~mag (3$\sigma$ confidence). At these depths, we would have detected all known long and short GRB host galaxies at $z<1.4$ \cite{Leibler+10,Lyman+17}. An underlying, unseen dwarf galaxy of G2 (Figure~\ref{fig:Hubble}) hosting a dust-obscured supernova is also difficult to reconcile, as low-mass galaxies have, in general, lower dust contents (see Methods section `Alternative Interpretations'). We conclude that a kilonova is the most natural explanation of the known channels to reproduce our observations of GRB\,211211A.

The interpretation of the near-IR excess as an $r$-process kilonova in turn implies that GRB\,211211A originated in a compact object merger. We briefly explore several explanations for extended gamma-ray emission following such an event. First, the extended emission may be explained by a relativistic wind imparted by a magnetar remnant \cite[e.g., ][]{Metzger+08_magnetarEE}. The progenitor may also have been a NS-black hole (BH) system. Tidal disruption of the NS would cause additional mass to fall back onto the remnant for several seconds following the merger and be launched in the jet, producing extended emission \cite[e.g., ][]{Metzger+10_rproc_fb,Desai+19}. However, we note that the moderate-sized blue component of the kilonova is not consistent with such a scenario. A BNS merger with a significantly asymmetric mass ratio provides a similar but alternate explanation, but may also struggle to produce sufficient blue ejecta. Future detections of GRBs and kilonovae in tandem with inferred properties from GW observations, which provide insight to the progenitor system's total and component masses, will elucidate the source of gamma-ray extended emission.
The detection of a kilonova following a long GRB implies that the current NS merger rates calculated from short GRBs (e.g., \cite{fong+15}) may underestimate the true population.

GRB\,211211A lies at a luminosity distance of 350 Mpc. This distance is only slightly beyond the sky and orientation-averaged horizon for the LIGO/Virgo detectors at design sensitivity. Notably, sensitivity is maximized for face-on mergers (i.e. events with GRBs pointed in our direction). Using GW template waveforms and expected noise curves (see Methods section `Gravitational Wave Detection Significance'), we calculate the expected signal-to-noise ratio (S/N) for a 1.4+1.4 M$_{\odot}$ binary merger at 350~Mpc during the third (O3), fourth (O4) and fifth (O5) observing runs, finding S/N of 7.4, 11.9 and 18.9, respectively. The S/N is even higher in the case of a fiducial 1.4+5 M$_{\odot}$ NS-BH merger (and S/N $> 10$ in O3), demonstrating that a GRB\,211211A-like event would be detectable in upcoming observing runs. Indeed, since the time-coincidence of GW and GRB emission and the known sky location can be used to increase the sensitivity of the GW detectors, such long GRB/GW coincidences can increase the number of multi-messenger signals that can be recovered in the future.



\vspace{0.5in}

\noindent \textbf{Figure 1. \textit{Swift}/BAT and \textit{Fermi}/GBM gamma-ray light curves of GRB\,211211A.} The light curves show similarities with both long GRBs and EE-SGRBs. We separate the \textit{Swift}/BAT light curve by band (\textbf{a}; blue) and normalized by the maximum number of counts in each band. The red curve is the light curve across all four bands, and is also normalized by the maximum counts. The two initial spikes (lasting $\sim 4$ and $\sim 8$~s) are prominent in each of the bands shown, while the tail ($\gtrsim 12$~s) becomes softer over time. While this soft tail is similar to the behavior of past EE-SGRBs (e.g., GRB\,060614 \cite{Gehrels+06}), its initial pulses are longer than those previously observed in EE-SGRBs. The \textit{Fermi}-GBM light curve of GRB\,211211A shows a similar structure to that of BAT (\textbf{b}). We also show the hardness ratio (\textbf{c}; the ratio of 50 -- 300\,keV to 10 -- 50\,keV photon fluxes) versus $t_{90}$ for GRBs in the \emph{Fermi}-GBM GRB catalog \cite{vonKienlin20}. The $t_{90}$ time-averaged properties of GRB\,211211A (blue) are typical of long GRBs, which occupy the lower-right corner of the parameter space.

\noindent \textbf{Figure 2. The field of GRB\,211211A in {\em Hubble Space Telescope} ({\em HST}) and Gemini-North imaging}. Late-time {\em HST} F606W and F140W images (\textbf{a,c}, respectively) covering the position of the {\em Swift}/XRT afterglow (blue circle) and the NIR counterpart (red crosshairs). We label the putative host, SDSS J140910.47+275320.8 ($z=0.076)$, which is offset $5.\!\!''44$ from the NIR counterpart and a second nearby galaxy (``G2''; see Methods section `Strong Evidence in Favor of a $z=0.076$ Origin'). No source is detected at the position of the NIR counterpart to a depth of $\mathrm{F606W} > 27.8$~mag. A smoothed Gemini/NIRI $K$-band image at 4.1 days post-burst detects a $K = 22.4$~mag point source at the position of GRB\,211211A's optical afterglow (\textbf{b}).

\noindent \textbf{Figure 3. Afterglow and kilonova models fit to selected observations of GRB\,211211A's broadband counterpart.} Modeling strongly supports the detection of an $r$-process enriched component. In (\textbf{a}) we plot relevant detections and their 1$\sigma$ uncertainties, 3$\sigma$ upper limits alongside the superimposed kilonova and afterglow models (solid lines) and the afterglow model alone (dotted lines). The afterglow model is well-constrained by the radio and X-ray light curves and provides a good fit to the optical data at $\lesssim 0.1$~day post-burst. The NIR detections are $\approx 4$~magnitudes brighter than predicted by the afterglow model and require a kilonova component to fit. In (\textbf{b}) the kilonova model (solid lines) provides a reasonable fit to the afterglow-subtracted optical-NIR light curve. We also plot models tuned to AT\,2017gfo from \cite{Nicholl+21} shifted to the redshift of GRB\,211211A (dashed lines). The $K$-band light curves are approximately the same luminosity at 4.1 days post-burst and fade on similar timescales.

\section*{Methods}
\label{sec:methods}

Unless otherwise stated, we report all observations in AB mag units and all times in the observer's frame. We use a standard cosmology of $H_{0}$ = 69.6~km~s$^{-1}$~Mpc$^{-1}$, $\Omega_{M}$ = 0.286, $\Omega_{vac}$ = 0.714 throughout this work \citep{Bennett+14}.

\bmhead{Gamma-ray Burst Detection}

The refined {\em Swift}/BAT position localizes GRB\,211211A to R.A. = \ra{14}{09}{05.2}, decl. = \dec{+27}{53}{03.8} with an uncertainty of $1'$ \cite{GRB211211a_bat}. GRB\,211211A was also identified by the Gamma-ray Burst Monitor (GBM) on board the \textit{Fermi Space Telescope} with a consistent localization \cite{Meegan+09,grb211211a_gbm}. The burst was further detected by the CALET Gamma-ray Burst Monitor \cite{CALET,grb211211a_CALET} and the INTEGRAL SPI-ACS \cite{INTEGRAL_SPI,grb211211a_integral}. 

\bmhead{\textit{Swift} Observations}

The \textit{Swift} X-ray Telescope (XRT; \cite{Swift-XRT}) observed the field of GRB\,211211A from $\delta t =$ 69 s to 74.2 ks (where $\delta t$ is time since the BAT trigger), identifying an uncatalogued X-ray source at a refined position of R.A. = \ra{14}{09}{10.08}, decl. = \dec{+27}{53}{18.8} with an uncertainty of $1.9''$.

X-ray data are downloaded from the UK \emph{Swift} Science Data Centre \cite[UKSSDC;][]{Evans07,Evans+09}. We take the $0.3$ -- $10$\,keV flux light curve and convert it to 1\,keV flux density \cite[cf.][]{Gehrels08} using the photon index of $1.51$ from the late time-averaged photon counting spectrum on the UKSSDC \cite{Evans07,Evans+09}. The early X-ray light curve (taken in windowed timing mode) shows a bright plateau with a $0.3$ -- $10$\,keV flux of $\sim 3\times10^{-8}$\,erg\,s$^{-1}$\,cm$^{-2}$. Its subsequent rapid decay at a rate of $\gg t^{-3}$ indicates an internal origin for the emission. Photon counting mode data taken from several thousand seconds after trigger show a shallower power law evolution, consistent with the emergence of the afterglow.

The {\it Swift}/UVOT began settled observations of the field of GRB 211211A 88 s after the BAT trigger. The afterglow was detected in all of the UVOT filters. To reduce contamination from the nearby galaxy, source counts were extracted from the UVOT image mode data using a source region of $3''$  radius. In order to be consistent with the UVOT calibration, these count rates were then corrected to $5''$ using the curve of growth contained in the calibration files. Background counts were extracted using a circular region of radius $20''$ located in a source-free region near to the GRB. The count rates were obtained from the image lists using the {\it Swift} tool \texttt{uvotsource}. They were converted to magnitudes using the UVOT photometric zero points \cite{poole+08,breveld+11}. To improve the signal-to noise ratio, the count rates in each filter were binned using $\Delta t = 0.2\delta t$. We report all UVOT photometry in Extended Data Table~\ref{tab:swift_obs}. The detection of the afterglow in six \textit{Swift}/UVOT filters strongly supports a $z\lesssim1.4$ origin for GRB\,211211A (99\% confidence level).

\bmhead{Radio Observation}

We initiated 6~GHz ($C$-band) VLA observations of GRB\,211211A at $\delta t = 6.27$~days (Program \#21B-198; PI: Fong). We used 3C286 for flux and bandpass calibration and J1407+2827 for gain calibration. We employed the Common Astronomer Software Application (CASA) pipeline products for data calibration and analysis \cite{CASA}, and imaged the source using CASA/\texttt{tclean}, using a Briggs weighting and robustness parameter of 0.5.  No source is detected at the position of the X-ray afterglow to a 3$\sigma$ (5$\sigma$) upper limit of 9.6~$\mu$Jy (16~$\mu$Jy). We utilize the more conservative, 5$\sigma$ upper limit in our analysis to account for any effects from scintillation.

\bmhead{Optical Afterglow Observations}

At 16.6~hr post-burst, we obtained $gri$-band imaging and spectroscopy of the GRB counterpart and putative host galaxy \cite{GCN31221}, using the Alhambra Faint Object Spectrograph and Camera (ALFOSC) mounted on the 2.6m Nordic Optical Telescope (NOT). We reduce the images using standard techniques and find that the afterglow is well detected in all filters. We flux calibrate images using standard stars in the field from the Pan-STARRS catalog \cite{Chambers+16}. We obtained spectroscopy using grism \#4, which covers the wavelength range 3500--9500~\AA{} at resolution $\lambda/\Delta{\lambda} = 350$. We oriented the slit to cover both the counterpart and the nearby galaxy. We detect a featureless continuum from the transient, hampering a direct redshift measurement (Extended Data Figure~1).

We obtained four epochs of $i$-band imaging with the Calar Alto Faint Object Spectrograph mounted on the 2.2m CAHA Telescope over $\delta t = 0.7 - 19.6$~days (Program 21B-2.2-018 PI: de Ugarte Postigo). We reduce images following standard procedures in \texttt{IRAF}. We perform aperture photometry on the images with IRAF/\texttt{phot} (\cite{Tody93}).

\bmhead{Further Optical-NIR Observations}

We initiated NIR observations with the Near-Infrared Imager (NIRI; \cite{NIRI03}) mounted on the 8m Gemini-North telescope (Program GN2021B-Q-109; PI: Fong) on 2021 December 15 ($\delta t = 4.1$~days). We detect a $K$-band source at R.A. = \ra{14}{09}{10.119}, decl. = \dec{+27}{53}{18.06} (error of $0.\!\!''19$), consistent with the X-ray and optical afterglow positions. We continued to observe $\sim$nightly in the $i$-, $J$- and/or $K$-bands with NIRI and the Gemini Multiple Object Spectrograph (GMOS; \cite{GMOS04}) on Gemini-North and the MMT and Magellan Infrared Spectrograph (MMIRS) mounted on the 6.5m MMT (\cite{MMIRS12}, Programs UAO-G178-21B, UAO-S127-21B; PIs: Rastinejad, Smith) until 2021 December 21 ($\delta t = 9.98$~days).

We reduced NIRI images using the Gemini DRAGONS pipeline \cite{DRAGONS19}, and GMOS and MMIRS images with a custom python pipeline, {\tt POTPyRI}\cite{POTPyRI}. Images were astrometrically registered to SDSS or the Gaia catalog using standard IRAF tasks, the \texttt{Gaia} software or \texttt{astrometry.net} \cite{lang10}. Between $\delta t = 4.1$ and $8.0$~days, our NIR observations clearly detect a source at the position of the optical afterglow. To ensure no host galaxy flux is contaminating our photometric values, we obtained a deep $i$-band template image of the field at $\delta t \approx 55$~days with Gemini-N/GMOS. Further, we obtained deep $K$ and $K_{s}$-band template images of the field at $\delta t \approx 66, 88$ and $98$~days with the Espectr\'ografo Multiobjeto Infra-Rojo (EMIR) mounted on the 10.4m Gran Telescopio Canarias (GTC, Program GTCMULTIPLE2H-21B; PI: de Ugarte Postigo), MMT/MMIRS and the LBT Near Infrared Spectroscopic Utility with Camera and Integral Field Unit for Extragalactic Research (LUCI; \cite{LUCI}) mounted on the dual 8.4m mirrored Large Binocular Telescopes (LBT, Program IT-2021B-018; PI: Palazzi), respectively. We reduce the EMIR data using a self-designed pipeline based on shell scripts and \texttt{IRAF} tasks and the LBT $K_s$-band image using the data reduction pipeline developed at Osservatorio Astronomico di Roma (INAF; \cite{Fontana2014a}).

We aligned the images using standard IRAF tasks and perform image subtractions using \texttt{HOTPANTS} \cite{becker15}. For the $i$-band image at $\delta t = 5.1$~days, we clearly detect a residual in the subtraction. Due to the faintness of the NIR detections and the added noise of image subtraction, no source is detected in the $K$-band residuals. However, our template image allows us to place limits on any underlying source contribution to $K_s > 24.6$~mag (3$\sigma$).

We calibrate the Gemini-N/GMOS, MMT/MMIRS, GTC/EMIR and LBT/LUCI images using stars in common with Sloan Digital Sky Survey Data Release 12 (SDSS DR12; \cite{Alam+15}) and 2-Micron All Sky Survey (2MASS; \cite{2MASS}). Due to NIRI's narrow field of view and the resulting dearth of 2MASS standard stars, we calibrate NIRI images using stars in common with the MMIRS image taken at $\delta t = 7.98$~days whose magnitudes we have measured from comparison to 2MASS. We perform aperture photometry at the position of the afterglow using the IRAF/\texttt{phot} task on the $i$-band subtracted image and the $J$- and $K$-band images directly. We derive upper limits on the Gemini, LBT and MMT images by measuring the magnitudes of 3$\sigma$ sources in the field using an aperture approximately proportional to the full width-half maximum (FWHM) of the transient.

Finally, at 17.6~days post-burst, we obtained $i$-band imaging of the field with the NOT/ALFOSC. We do not detect a source at the position of the optical/NIR counterpart to a 3$\sigma$ limiting magnitude of $i>24.7$~AB mag. We report all photometry in Extended Data Table~\ref{tab:observations} and plot spectral energy distributions (SEDs) of the UV-optical-NIR counterpart to GRB\,211211A at 5 approximately contemporaneous epochs in Extended Data Figure~2.

\bmhead{Hubble Space Telescope Observations}
\label{sec:host_obs}

On 2022 April 12 and 14 we observed the field of GRB\,211211A using {\em HST} with WFC3/IR/F140W and ACS/WFC/F606W, respectively (Program 16923; PI: Rastinejad). We reduced the images using the custom pipeline \texttt{hst123}\cite{hst123}, which uses the \texttt{astrodrizzle} package to reduce and align the images \cite[][ for details]{Kilpatrick+22_GW170817}. We performed aperture photometry on SDSS J140910.47+275320.8 using a 5'' aperture in the drizzled images and using zero points calculated for the drizzled frames by {\tt hst123}. After aligning our {\it HST} data to our $K$-band Gemini image from $\delta t = 4.1$~days, we place upper limits on the presence of an {\it HST} counterpart using artificial star injection at the NIR counterpart position. Using the fake star methods in {\tt dolphot} \cite{dolphot}, we inject 50,000 artificial stars in increments of 0.01~mag.  We then estimate the magnitude threshold at which 99.7\% of sources are recovered at 3$\sigma$, which we consider to be the 3$\sigma$ limiting magnitude. No source is detected at the position of the optical afterglow to $\mathrm{F606W} > 27.76$~mag and $\mathrm{F140W} > 27.19$~AB mag. We list additional multi-band limits for an underlying source from ground-based telescopes in Extended Data Table~\ref{tab:observations}.

Our \textit{HST} upper limits eliminate the presence of an underlying host with the brightness of any known short and long GRB hosts at $z < 1.4$ \cite{Leibler+10,Lyman+17} (where $z<1.4$ is the upper limit from the UVOT afterglow detections) as well as a galaxy of $\gtrsim 0.01$ L$^*$ at $z < 1.4$ (where L$^*$ is the characteristic galaxy luminosity parameter; \cite{Brown+01,Wolf+03,Willmer+06,ReddySteidel09,Finkelstein+15}). The {\em HST} limit ($M_{\rm F606W} \gtrsim -9.9$~mag) also allow us to rule out $\approx 46$\% of the dwarf galaxies that might be associated with SDSS J140910.47+275320.8 (based on Local Group dwarf galaxy luminosities; \cite{McConnachie12}).

\bmhead{Strong Evidence in Favor of a $z=0.076$ Origin}

Using the Gemini/NIRI image at $\delta t = 4.1$~days (Figure~\ref{fig:Hubble}), we measure an offset of $5.\!\!''44\pm 0.\!\!''02$~(7.91 $\pm$ 0.03~kpc at $z=0.076$) between the center of the host galaxy and the position of the optical afterglow. This is within the range of expected offsets from both short and long GRBs, but more consistent with the range for short bursts \cite{FongBerger13,Blanchard+16,Lyman+17,Fong+22}. 

We note the presence of a second galaxy to the Northwest of SDSS J140910.47+275320.8 (``G2'' or SDSSJ140909.60+275325.8; Figure~\ref{fig:Hubble}). Using the Gemini/NIRI image at $\delta t = 4.1$~days we measure an offset between the NIR source and ``G2'' of $10.\!\!''30 \pm 0.\!\!''02$. Assuming the spectroscopic redshift  for G2 reported in SDSS DR12 \cite{Alam+15}, $z=0.4587$, this is a physical offset of 60.55 $\pm$ 0.12~kpc. We measure $r_{\rm G2} = 20.80 \pm 0.05$ mag from the Binospec template image (see below), and calculate a value of P$_{cc, {\rm G2}} = 13.3$\%. At the spectroscopic redshift of G2, GRB\,211211A's counterpart's peak $K$-band luminosity is greater than that predicted by kilonova models ($\nu L_{\nu} =  4 \times 10^{42}$~erg~s$^{-1}$). The relatively large P$_{cc, {\rm G2}}$ and greater projected and physical offsets in comparison to those of SDSS J140910.47+275320.8 strongly disfavor the potential association between G2 and GRB\,211211A.

\bmhead{Host Galaxy Observations}

On 2022 January 27, we obtained further optical observations in the $grz$-bands with the Binospec instrument mounted on the MMT (\cite{Binospec+19}; Program UAO-G178-21B; PI: Rastinejad). We calibrate images to SDSS DR12 and perform aperture photometry on SDSS J140910.47+275320.8 with IRAF/\texttt{phot}. We obtain further host photometry from template observations and the \textit{HST} images. We retrieve $u$-band photometry of the host from the SDSS archive \cite{Alam+15} and W1 photometry from WISE \cite{WISE_Wright+10}. We obtain UV photometry from \textit{Swift}/UVOT ($v,b,u,uvw1,uvm2,uvw2$-bands). We list all host photometry in Extended Data Tables~\ref{tab:observations} and \ref{tab:swift_obs}.

We obtained additional spectroscopy of SDSS J140910.47+275320.8 with the DEep Imaging Multi-Object Spectrograph (DEIMOS) mounted on the 10m Keck II Telescope on 2022 January 8 ($2\times1500$~s; Program O300; PI: Blanchard). The spectrum was observed with a 1''~slit and the 600ZD disperser at a central wavelength of 6500\AA with the GG455 blocking filter, covering the wavelength range $\approx$4500--9000\AA. We apply an overscan subtraction, flat-field corrections, model the sky background, and remove cosmic rays using \texttt{PypeIt} \cite{pypeit}. We also apply a wavelength calibration with KrXeArNeCdZnHg arc lamp spectra. Using \texttt{PypeIt}'s boxcar method with a 1.5'' radius to encapsulate the entire galaxy's light, we extract the 1D spectrum from both target science frames. We flux calibrate the spectra with the standard star HZ44, taken the same night as the science target, and co-add the 1D galaxy spectra. Finally, we apply a Galactic extinction correction in the direction of the target using the model of \cite{ccg+1989} and $A_{\rm V, ext}$ from the dust extinction maps of \cite{SchlaflyFinkbeiner11}. We confirm the redshift of $z=0.0763 \pm 0.0002$ ($347.8^{+1.0}_{-0.9}$~Mpc) from the identification of the H$\alpha$,  H$\beta$, [OIII]$\lambda 4958,5007$, [NII]$\lambda 6549,6584$ and [SII]$\lambda 6717, 6731$ emission lines.

\bmhead{Stellar Population Modeling of SDSS J140910.47+275320.8}
\label{sec:Prospector}

We model the stellar population properties of SDSS J140910.47+275320.8 using \texttt{Prospector}, a Python-based stellar population inference code \cite{Leja2019, jlc+2021}. We determine properties such as the total mass formed ($M_F$), age of the galaxy at the time of observation ($t_{\textrm{age}}$), optical depth, star formation history (SFH), and stellar ($Z_*$) and gas-phase ($Z_\textrm{gas}$) metallicities from jointly fitting the photometric and Keck/DEIMOS spectroscopic data at the galaxy's redshift. We apply a nested sampling fitting routine with \texttt{dynesty} \cite{Dynesty} to fully sample the parameter space of each property and build model SEDs using \texttt{FSPS} (Flexible Stellar population synthesis) and \texttt{Python-fsps} \cite{FSPS_2009, FSPS_2010}. Within the \texttt{Prospector} fit, we use the Milky Way Extinction Law \cite{ccg+1989} and assume a Chabrier initial mass function (IMF; \cite{Chabrier2003}). We apply a parametric delayed-$\tau$ SFH ($\text{SFH} \propto t \times e^{-t/\tau}$), characterized by the $e$-folding time $\tau$, which is a sampled parameter in the \texttt{Prospector} fit. We include the Gallazzi mass-metallicity \cite{gcb+05} relation to ensure that \texttt{Prospector} only samples realistic $M_F$ and $Z_*$ values and enforce a $2:1$ dust ratio between old and young stellar populations, as younger stars are observed to attenuate dust twice as much as old stars \cite{cab+20, Leja2019}. We build the model spectral continuum from a $10^{\rm th}$-order Chebyshev polynomial and model spectral line strengths and widths with a nebular emission model, which includes a gas ionization parameter and $Z_\textrm{gas}$. We further apply a noise inflation model to the observed spectrum to ensure proper weighting of the photometry against the high signal-to-noise spectrum. Finally, we convert $M_F$ to a stellar mass ($M_*$), $t_\mathrm{age}$ to a mass-weighted age ($t_m$), and the optical depth to $V$-band magnitude ($A_V$) using the equations in \cite{Nugent+20}.

We find that SDSS J140910.47+275320.8 has $t_m = 4.00^{+0.65}_{-0.59}$~Gyr, $\log(M_*/M_\odot) = 8.84^{+0.10}_{-0.05}$, $A_V = 0.05^{+0.04}_{-0.03} $~mag, $\log(Z_*/Z_\odot) = -0.69^{+0.09}_{-0.20}$, and log(Z$_\textrm{gas}$/Z$_\odot$) $= 0.22^{+0.77}_{-0.34}$. We show the \texttt{Prospector} SED fit compared to the observed data in Extended Data Figure 1. We determine an SED star formation rate (SFR) using Equation (1) in \cite{Nugent+20} and find the galaxy has a low SED-inferred SFR $=0.07$~M$_\odot$yr$^{-1}$ and specific SFR (sSFR) $\approx 0.10$~Gyr$^{-1}$. Following the methods in \cite{Kennicut1998} and \cite{Moustakas2006}, we also determine an SFR from the model-predicted emission line flux of H$\alpha$, finding SFR = $0.76 \pm 0.01~M_\odot$~yr$^{-1}$, which is higher than the SED-inferred SFR. We note that SED-inferred SFRs are typically systematically lower than H$\alpha$-inferred SFRs \cite{Leja2019}. From Equation (2) in \cite{Tachella2021} and using the SED-inferred sSFR and redshift, we determine that the galaxy is star-forming. 

Compared to the population of short GRB hosts \cite{BRIGHT_II}, SDSS J140910.47+275320.8 lies in the bottom $\approx 11.8\%$ of stellar masses, $\approx 86.7\%$ for stellar population age, and $19.1 \%$ inferred SFR. We note that it has much less star formation for its given stellar mass than other short GRB hosts, and is the lowest redshift star-forming host compared to the population \cite{BRIGHT_II}. Furthermore, SDSS J140910.47+275320.8 has distinct properties from NGC4993, the quiescent host of GW/GRB170817 \cite{Abbott+17a,bbf+17,Levan+17}. NGC4993 is $\approx 9$~Gyr older, $10^2$ times more massive, and has much less ongoing star formation ($\approx 10^{-4}$~Gyr$^{-1}$; \cite{bbf+17, BRIGHT_II}). Despite these contrasts and considering the low amount of active star formation in the host and its old stellar population age, we find little evidence from the host galaxy that GRB\,211211A originated in a young massive stellar progenitor.

\bmhead{Afterglow Model}
\label{sec:ag_modeling}

To model the synchrotron afterglow, we employ the methods of \cite[][ and references therein]{Lamb+17,Lamb+18,Lamb+21} and calculate the dynamics of a relativistic blast wave with the analytical solution of \cite{Pe'er12}. This solution assumes a uniform interstellar medium (ISM) environment, which is consistent with our results from spectral fitting. We do not find evidence for a reverse shock in the afterglow observations, and thus model only a forward shock. The eight physical parameters in our model are the inclination between the line-of-sight and the jet's central axis ($\iota$), isotropic equivalent jet kinetic energy ($E_{\rm k,iso}$), the electron distribution index ($p$), the jet half opening angle ($\theta_{\rm c}$), the Lorentz factor ($\Gamma$), the circumburst environment density ($n$), the fraction of energy that goes into the magnetic field ($\varepsilon_{B}$) and the electrons ($\varepsilon_{E}$). Our model solves for the order of the synchrotron break frequencies due to synchrotron self-absorption ($\nu_{\rm a}$), electron cooling ($\nu_{\rm c}$) and the minimum Lorentz factor in the distribution of shocked electrons ($\nu_{\rm m}$). We constrain the value for $p$ based on fits to the X-ray data.

We use \texttt{emcee} \cite{emcee} to determine a best-fit afterglow model and posterior distributions for the physical parameters. We fit our model to the entire X-ray and radio datasets and to the Galactic extinction-corrected UV-optical data at $\delta t < 0.1$~day, when the synchrotron afterglow luminosity is expected to dominate the kilonova contribution. Overall, our model provides a good fit to the observed data (Figure~\ref{fig:lc_models}). We find a slow-cooling spectrum with $\nu_{\rm a} < \nu_{\rm m} < \nu_{\rm c}$, where $\nu_{\rm c}$ is above the X-ray frequency, and $\nu_m$ below the UV-optical. 
In Extended Data Table~\ref{tab:ag_params}, we present the median and 1$\sigma$ errors of the physical parameters found by our best-fit model. Early UVOT data constrains the Lorentz factor within our afterglow model to $\Gamma \approx 70$, consistent with prompt emission analysis of the burst \cite{Gompertz+22}.

Both our model parameters and the observations are in keeping with those seen in the short GRB population \cite{fong+15}, including GRB\,170817A (e.g., \cite{lamb+19}). The X-ray luminosity of the afterglow on timescales $>1000$ s lies roughly at the median of short GRB afterglows, while the earlier X-ray data is consistent with short GRBs with extended emission \cite{Gompertz+22}. The optical afterglow is also consistent with short GRBs \cite{fong+15}. At later times ($>2$ days) our extrapolation is fainter than afterglow detections of some bursts. However, upper limits of numerous other bursts are available at this epoch.

\bmhead{Kilonova Model}
\label{sec:kn_modeling}

We isolate the KN light curve by subtracting the median afterglow model from the optical and infrared data, propagating the 1-sigma uncertainties in the afterglow luminosity for each observation into the subtracted data. We fit this afterglow-subtracted photometry to a suite of kilonova models using the prescriptions of \cite{Cowperthwaite+17,villar+17,Nicholl+21} within the Modular Open Source Fitter for Transients (\texttt{MOSFiT}; \cite{Guillochon+18}). The luminosity in these models is produced by the radioactive decay of $r$-process elements, and diffuses out of the ejecta following the standard formalism given by \cite{Arnett82}. The ejecta in our models consists of three components produced by different processes in the merger, and each has a separate mass, velocity and composition, with more lanthanide-rich material (arising in regions of lower electron fraction and/or neutrino irradiation) having a higher opacity. While the afterglow model is only fit to $\gtrsim 3 \sigma$ UVOT detections, the kilonova model is fit to UVOT detections at the $\gtrsim 1 \sigma$ level, providing information on the contribution of the shocked cocoon. In Extended Data Table~\ref{tab:swift_obs} we separately list the photometry used in the afterglow and kilonova modeling.

Interactions between the compact object progenitors produce dynamical ejecta just prior to and during the merger. ``Blue'' ($\kappa = 0.5 $~cm$^{2}$\,g$^{-1}$) material is ejected in the polar direction and assumed to be lanthanide-free due to strong neutrino irradiation, either due to the contact shock or surface winds from a magnetar remnant (hence this component is unlikely to exist in a NS-BH merger). Interaction from the jet may also lower the lanthanide fraction of material ejected at the poles \cite{Nativi+21}. Conversely, ``red'' ($\kappa = 10$~cm$^{2}$\,g$^{-1}$) dynamical ejecta is produced by tidal tails and is concentrated along the equatorial axis. A post-merger accretion disk formed around the remnant object provides a second source of kilonova ejecta. The amount of material ejected by the disk is dependent on the merger remnant (e.g., a prompt-collapse BH or a short-lived NS; \cite[e.g., ][]{metzgerfernandez14}), and can vary by orders of magnitude (in terms of $M_{\odot}$).  The opacity (i.e.~composition) depends on the exposure to neutrino flux, thought to be higher for a longer-lived NS remnant \cite[e.g., ][]{Lippuner+17}. Light curve models for AT\,2017gfo suggested that this component had an intermediate ``purple'' opacity ($\kappa \approx 3$~cm$^{2}$\,g$^{-1}$) \cite{villar+17}.

The relative contribution to the total luminosity by each spatially distinct component depends on the observer viewing angle \cite{Darbha2020}. Given the bright on-axis GRB, we assume a viewing angle along the binary's polar axis. The luminosity of blue ejecta can be further enhanced by shock heating from the GRB jet traversing the ejecta (e.g., \cite{Kasliwal+17,Arcavi18}), which we include in our model following \cite{Piro+18}. We modify their prescriptions by adding a constraint that shock cooling ceases to contribute luminosity once the cocoon becomes optically thin ($\lesssim 1$~day; equation 14 in \cite{Piro+18}). Our models do not include the effects of jet interaction (e.g., \cite{Nativi+21}) or magnetic fields \cite{Metzger+18,Radice+18,CiolfiKalinani20}.

We fit the data using two variations of this model. We adopt flat priors on all parameters in both cases, and use \textsc{dynesty} \cite{Speagle2020} to sample the posteriors. We include a white noise parameter, $\sigma$, in the likelihood function as in \cite{Guillochon+18}. First we use a model based on \cite{villar+17} and let the mass $M_{\rm ej,i}$ and velocity $v_{\rm ej,i}$ of each ejecta component vary freely. We also include the effects of (fixed) viewing angle, and allow the fraction of blue ejecta in the shocked cocoon ($\zeta_{\rm shock}$) to vary, both following \cite{Nicholl+21}. This model produces the best-fit light curve in Figure \ref{fig:lc_models}. While the model provides a good fit to the NIR points, it over-predicts the $i$-band luminosity for the two detections at $\delta t \gtrsim 2.5$~days. We note that both of these points have high systematic (precise flux measurements vary up to 1~mag with aperture choice) and statistical (Extended Data Table~\ref{tab:observations}) errors. The posterior distributions of the model parameters are shown in Extended Data Figure 3. The total model evidence returned by \textsc{dynesty} is $\ln(Z)=24.9$. Derived ejecta masses and velocities are overall similar to inferences for GW170817 \cite{villar+17}. The main difference is in the ratio of red to purple ejecta, with a larger red mass preferred in GRB\,211211A due to the redder $J-K$ and $i-K$ colors at $\sim 1$ week post-merger.

Although the statistical errors shown in Extended Data Figure 3 are generally $\lesssim10\%$, the model assumptions of constant grey opacities for each component likely implies a non-negligible systematic error. The opacity is degenerate with ejecta mass and velocity through the light curve diffusion timescale, $\tau \propto(\kappa M/v)^{1/2}$, implying an additional fractional uncertainty on the $r$-process yield up to $dM/M \sim d\kappa/\kappa \sim 1$. However, the true systematic error is lower than this because $M$ is also directly tied to the radioactive heating rate, and experiments with freeing the opacities suggest it is $\sim 50\%$. In addition, we attempted to fit the data with a two-component model, allowing the opacity of the redder component to vary. In this case, we still recover a total $r$-process mass of $\sim 0.05$ M$_\odot$, though the derived opacity (2\,cm$^2$\,g$^{-1}$) and velocity ($>0.3c$) do not naturally align with an expected ejecta component (see `Binary-Based Kilonova Model').

To determine if there are any detectable degeneracies between the afterglow and kilonova posteriors, we perform an approximate joint fit to the data. Adding the kilonova light curves to those of the afterglow during the inference process requires fitting a 15-parameter model. Thus, the MCMC samplers naturally struggle to find the global optimum and do not reach convergence in our tests. However, the joint fit does not detect any degeneracies, indicating that the effect of the uncertain afterglow model flux on the kilonova is negligible. The flux contrast between the kilonova and afterglow light curves also supports this: during the time of the kilonova detections, most epochs have kilonova fluxes 1-2 orders of magnitude above those of the afterglow. Thus, changes in the afterglow within model uncertainties affect the kilonova only at the few-percent level. At earlier times, when the kilonova is not clearly visible above the afterglow, we have added an earlier \textit{Swift}-UVOT unfiltered observation to better constrain the early light curve. We find that the resultant changes in the afterglow parameters are small, producing a moderately lower Lorentz factor with other parameters largely unchanged. This suggests that the afterglow model is quite robust, and that the kilonova is not sensitive to allowed changes in the afterglow. 

Since we can find no evidence for strong degeneracies in the models, we have employed our two-step approach using the well-tested codes optimized individually for the afterglow and kilonova \cite{Lamb+21,Nicholl+21}. We propagate all uncertainties in the optical afterglow light curves into the subtracted kilonova photometry before fitting. We also report the afterglow-subtracted photometry in Extended Data Tables~\ref{tab:observations} and \ref{tab:swift_obs}.

\bmhead{Binary-Based Kilonova Model}
Advancements in the theoretical modeling of compact object mergers and their outflows have made it possible to tie kilonova observations to properties of the progenitors and remnant (e.g., \cite{Coughlin+19_BNSparam, Nicholl+21}). Specifically, the masses of the dynamically-ejected, lanthanide-rich red and lanthanide-poor blue components are determined by the progenitor mass ratio ($q$; \cite{sekiguchi+16}), chirp mass ($\mathcal{M}$), and NS radius ($R_{\rm NS}$; e.g., \cite{DietrichUjevic17}). Similarly, an estimate of the intermediate opacity purple mass ejected by the post-merger accretion disk informs estimates of $\mathcal{M}$, $R_{\rm NS}$ and the NS remnant lifetime (e.g., \cite{Radice+18}). The ejecta velocities of each component further depend on $\mathcal{M}$ \cite{metzgerfernandez14}. Re-formulating the model in terms of pre-merger binary parameters allows us greater insight to the progenitor system, and ensures that $M_{\rm ej} - V_{\rm ej}$ combinations (and thus the resulting light curves) are consistent and realistic in the context of theoretical simulations.

We therefore fit the afterglow-subtracted photometry with the binary-based model of \cite{Nicholl+21}. We fix the viewing angle to pole-on, and the equation-of-state dependent parameters to the best-fit values for GW170817: $R_{\rm NS}=11.1$\,km and maximum stable mass $M_{\rm TOV}=2.17$\,M$_\odot$. The free parameters are the chirp mass, $\mathcal{M}=(M_1 M_2)^{3/5}(M_1+M_2)^{-1/5}$, and mass ratio, $q=M_2/M_1 \le 1$, where $M_1$ and $M_2$ are the masses of the two neutron stars, the fraction of the remnant disk ejected, the fraction of blue ejecta enhanced by NS surface winds for long-lived remnants, and the fraction of blue ejecta shocked by the GRB. We introduce one additional free parameter to the \cite{Nicholl+21} model: the time after merger at which the GRB jet re-heats the polar ejecta (a larger $t_{\rm shock}$ results in a brighter cocoon due to the larger radius of the ejecta). Even if the jet has already broken through the ejecta, recollimation shocks at the jet-ejecta interface may continue to appreciably heat ejecta material as long as the jet is active \cite[e.g.,][]{Gottlieb+20}, though the efficiency of such heating is likely to be lower than in the case of a choked jet \cite[e.g.,][]{Duffell+18}.  This additional freedom is motivated by the temporally extended GRB duration compared to GRB\,170817, and is required to match the early UV emission. The best-fit model is shown in Extended Data Figure~4, with posteriors shown in Extended Data Figure~5. The binary masses are $M_1=1.42\pm0.05$\,M$_\odot$ and $M_2=1.25\pm0.04$\,M$_\odot$, consistent with typical NSs (and indeed GW170817). The fraction of disk mass ejected is similar to the $\sim 0.1$ inferred for GW170817 by \cite{Nicholl+21}. We caution that the systematic errors in this model are also $\sim50\%$ \citep{DietrichUjevic17,Nicholl+21}. If a magnetar remnant is the source of the extended emission, we might expect a large value of the blue ejecta enhancement factor ($1/\alpha$) due to magnetic winds. The mode of the posterior is $\alpha=0.6$, but is not well constrained due to a degeneracy with $q$ visible in Extended Data Figure~5.

\bmhead{$^{56}$Ni-Powered Transient Model}
\label{sec:ni-model}
To further rule out any associated SN, or a white dwarf - NS merger \cite{King2007}, we also fit the light curve with a single-component model powered by $^{56}$Ni decay, using the default \textsc{mosfit} model. The free parameters in this case are the ejecta mass and velocity, the nickel fraction, the gamma-ray trapping efficiency, and a minimum (recombination) temperature. We fixed the optical opacity to $\kappa=0.2$~cm$^{2}$\,g$^{-1}$, appropriate for electron scattering for ionised intermediate mass or iron-group elements. This model is unable to provide a reasonable fit: the model evidence is $\ln(Z)=-59.6$ because it is too faint by several magnitudes during the first day (Extended Data Figure~6). Physically, the problem is that a single-component model cannot cool quickly enough to match both the early UV and late-time NIR light curves. The posteriors for velocity ($\approx 10^5$\,km\,s$^{-1}$) and nickel fraction ($\approx 1$) rail against the upper bounds of their priors. This model is therefore heavily disfavoured compared to the kilonova fits.

\bmhead{Alternative interpretations}
\label{sec:alternatives}

The optical-near-IR counterpart observed following GRB\,211211A is strongly reminiscent of the kilonova AT\,2017gfo, and, as our fitting shows, can be explained with a superimposed afterglow and kilonova model. Straightforward dust extinction in the host galaxy or burst vicinity cannot explain the counterpart's early blue color and its subsequent evolution from blue to red colors. The measured offset of GRB\,211211A's optical counterpart from the host galaxy center (7.91 $\pm$ 0.03~kpc) is highly consistent with the known offsets of short GRBs, which have a median of 7.92~kpc and span $\approx$1.79--28.63~kpc (16th and 84th percentiles, \cite{Fong+22}). GRB\,211211A's offset is less consistent but still within the range of known long GRB offsets, for which the median is 1.28~kpc and span $\approx0.075-14$~kpc \cite{Blanchard+16}). The counterpart fades much faster than the rate expected for more distant SN events (e.g. a dust obscured SN at $z > 0.5$). Comparing the $i$-band upper limit at $\delta t = 17.6$~days to the light curves of several long GRB SNe \cite{Galama+98,Matheson+03,Clocchiatti+11,Cano+14,Greiner+15,Cano+17} we find that none of the SNe are allowed by our upper limit out to $z=0.5$ ($M > -17.6$).

Recently, \cite{Waxman+22} suggested that GRB\,211211A's near-IR excess could instead be caused by an IR dust echo, a scenario in which dust local to the GRB (such as in a giant molecular cloud) is destroyed by the GRB jet directly along the line-of-sight and surrounding dust is heated and subsequently re-radiates. Direct light curve modeling is not straightforward for this scenario. Thus one cannot directly compare the dust model to that of a kilonova. We also note that such signatures have not been needed to explain previous LGRB afterglows, although there is a paucity of relevant IR searches in long GRBs.

The \cite{Waxman+22} scenario requires an underlying host fainter than that of all known GRBs at $z<3$ \cite{Lyman+17}. As the detection of the UVOT afterglow limits GRB\,211211A's origin to $z<1.4$, this explanation requires an extremely faint, low-mass host galaxy so-far unseen in the GRB host population \cite{Lyman+17,Fong+22,BRIGHT_II}. Since very low-mass galaxies are typically less dusty than more massive galaxies \cite{Santini+14,Calura+17}, this is not a probable location to observe a GRB within a dense, dusty environment. We also note that for a dusty line of sight we may expect to observe some residual absorption in the form of either excess $A_V$ or excess X-ray $N_H$. We find no evidence for either in the spectrum of GRB\,211211A.

One reason that \cite{Waxman+22} prefer the dust echo model over that of a kilonova is based on the observed kilonova color at 5.1 days. After afterglow subtraction the flux ratio at this epoch is red albeit with large uncertainties, $F_{K} / F_{i} = 43 \pm 29$. This differs from the colors of AT\,2017gfo at the same epoch at the $\sim 1.2 \sigma$ level. However, much redder colors at comparable luminosities can be found in other kilonova models, such as the lanthanide-rich models of \cite{Kasen+17} that were applied to AT\,2017gfo. We therefore do not believe that this color represents a problem for the kilonova interpretation.

Finally, the BAT light curves also show evidence for a short GRB-like origin. We cross-correlate BAT light curves covering the t$_{90}$ interval in 4 standard energy channels to measure the delay in the arrival times of soft photons compared to hard. In 1\,ms time bins, we find delays of $10 \pm 9$\,ms between 15 -- 25 and 50 -- 100\,keV photons (bands 1 and 3), and $4 \pm 9$\,ms between 25 -- 50 and 100 -- 150\,keV (bands 2 and 4). At $z = 0.076$, such small spectral lags are consistent with the distribution of short GRBs \cite{Bernardini15}, and inconsistent with the established long GRB lag-luminosity relationship \cite{Ukwatta10,Ukwatta12}. The expected peak luminosity from this relationship ($\gtrsim 10^{53}$\,erg\,s$^{-1}$) would require $z \sim 1.5$, an origin which is disfavored by our deep \textit{HST} observations and UVOT afterglow detection.

\bmhead{Gravitational Wave Detection Significance}
\label{sec:gw_observability}

To explore if the LIGO-Virgo network (H1, L1 and V1) could have detected the merger precipitating GRB\,211211A had it been operating at the time, we consider two representative cases for the progenitors: a $1.4+1.4M_{\odot}$ BNS merger, or a $1.4+5.0M_{\odot}$ NS-BH merger viewed face on (${\theta}_{jn} = 0$). Our calculations use a 2048 s duration data segment (chosen to be long enough even for a binary neutron star starting at 10~Hz) with a similarly high sampling frequency of 8192~Hz. We take the frequency integral between $f_{\rm low}$ = 10 Hz or 20 Hz and $f_{\rm high}$ = 4000~Hz and neglect component spins, orbital eccentricity and tidal effects. Although these parameters will affect the binary phasing, we expect them to have a very small effect on the S/N. We also neglect all cosmological effects and set the phase and polarizations angles to zero, as they will have negligible effect on the S/N. We utilize the waveform IMRPhenomPv2 NRTidal \cite{Hannam+14,Khan+16,Dietrich+17} called through \texttt{bilby} \cite{bilby,lalsuite} and obtain the noise power spectral densities (actual and predicted) from \url{https://dcc.ligo.org/LIGO-T2000012/public}.

For both the BNS and NSBH, we consider 4 representative cases: with $f_{\rm low}$ = 20 Hz, the O3 (actual), O4 and O5 (predicted) noise curves and O5 with a more optimistic low frequency cutoff of $f_{\rm low}$ = 10 Hz. We calculate all S/N using $D_L = 350$~Mpc, the time of the burst and the coordinates of GRB\,211211A's XRT position. We find that the BNS would not be detectable in O3 (S/N $\approx 7.4$), but the NSBH would have been (S/N $\approx 11.7$). The BNS and the NSBH would have had S/N $> 10$ in O4 and O5, likely making them detectable in GWs.

\bmhead{Comparison to AT\,2017gfo and Short GRB Kilonova Candidates}

Despite accompanying an event that is superlative in numerous ways, the kilonova of GRB\,211211A is unremarkable in luminosity and color compared to its few peers. In Extended Data Figure~7 we plot $i$ and $K$-band light curves of GRB\,211211A's kilonova along with AT\,2017gfo's light curve \cite{Andreoni+17,Arcavi+17,Coulter+17,Cowperthwaite+17,Diaz+17,Drout+17,Evans+17,Hu+17,Kasliwal+17,Lipunov+17, Pian+17, Pozanenko+17, Shappee+17, Smartt+17, Tanvir+17,Troja+17, Utsumi+17,Valenti+17,villar+17} and relevant rest-frame short GRB observations from the catalog of \cite{Rastinejad+21}. Due to the limits of current NIR detectors, we are only able to compare GRB\,211211A's rest-frame $K$-band light curve to that of AT\,2017gfo, though we include rest-frame $JH$-band short GRB kilonova observations for context (open symbols). At $\delta t \approx 5.1$~days, the only epoch of concurrent $i$- and $K$-band detections of GRB\,211211A, we measure a color of $(i-K) = 3.6$~mags. This is redder than the $(i-K) = 2.0$~mags measured at a similar rest-frame epoch for AT\,2017gfo \cite{villar+17}.

In Extended Data Figure~8, we plot the best-fit ejecta and mass velocity estimates for GRB\,211211A compared to those of AT\,2017gfo \cite[red boxes; compiled in ][ and references therein]{siegel19} and short GRB kilonova candidates \cite{barnes+16,lamb+19,troja+19}. Our estimates for GRB\,211211A are compatible with those of past kilonovae, including AT\,2017gfo. Estimates are highly model-dependent, and thus direct comparisons are not advisable.

\backmatter

\section*{Declarations}

\bmhead{Data Availability}

The majority of data generated or analysed during this study are included in this article's Extended Data Tables. Gamma-ray and X-ray light curves may be downloaded from the UK \emph{Swift} Science Data Centre and the online HEARSAC archive at \url{https://heasarc.gsfc.nasa.gov/W3Browse/fermi/fermigbrst.html}. Any additional data requests should be made to Jillian Rastinejad.

\bmhead{Code Availability}

The kilonova model scripts are available at \url{ https://github.com/guillochon/MOSFiT}. The scripts used to model the afterglow will be publicly available upon publication of this manuscript. The {\tt Prospector} stellar population modeling code is available at \url{https://github.com/bd-j/prospector}.

\bmhead{Acknowledgments}

We thank ShiAnne Kattner, Skyler Self, Joannah Hinz and Igor Chilingarian at the MMT and Jennifer Andrews and Kristin Chiboucas at Gemini Observatory for their assistance in obtaining observations. We thank Andreas von Kienlin for providing the GBM hardness versus duration data. We thank Patricia Schmidt and Geraint Pratten for assistance with the LIGO SNR calculations.

The Fong Group at Northwestern acknowledges support by the National Science Foundation under grant Nos. AST-1814782, AST-1909358 and CAREER grant No. AST-2047919. W. Fong gratefully acknowledges support by the David and Lucile Packard Foundation. 
A.J. Levan and D.B. Malesani are supported by the European Research Council (ERC) under the European Union’s Horizon 2020 research and innovation programme (grant agreement No.~725246).
M. Nicholl and B. Gompertz are supported by the European Research Council (ERC) under the European Union’s Horizon 2020 research and innovation programme (grant agreement No.~948381). M. Nicholl acknowledges a Turing Fellowship.
G. Lamb is supported by the UK Science Technology and Facilities Council grant, ST/S000453/1.
A. Rossi and E. Marini acknowledge support from the INAF research project ``LBT - Supporto Arizona Italia''. 
J. F. Ag\"u\'i Fern\'andez acknowledges support from the Spanish Ministerio de Ciencia, Innovaci\'on y Universidades through the grant PRE2018-086507. D. A. Kann and J. F. Ag\"u\'i Fern\'andez acknowledge support from Spanish National Research Project RTI2018-098104-J-I00 (GRBPhot).

W. M. Keck Observatory and MMT Observatory access was supported by Northwestern University and the Center for Interdisciplinary Exploration and Research in Astrophysics (CIERA). Some of the data presented herein were obtained at the W. M. Keck Observatory, which is operated as a scientific partnership among the California Institute of Technology, the University of California and the National Aeronautics and Space Administration. The Observatory was made possible by the generous financial support of the W. M. Keck Foundation. The authors wish to recognize and acknowledge the very significant cultural role and reverence that the summit of Maunakea has always had within the indigenous Hawaiian community. We are most fortunate to have the opportunity to conduct observations from this mountain. Observations reported here were obtained at the MMT Observatory, a joint facility of the University of Arizona and the Smithsonian Institution. 

Based on observations obtained at the international Gemini Observatory (Program ID GN2021B-Q-109), a program of NOIRLab, which is managed by the Association of Universities for Research in Astronomy (AURA) under a cooperative agreement with the National Science Foundation on behalf of the Gemini Observatory partnership: the National Science Foundation (United States), National Research Council (Canada), Agencia Nacional de Investigaci\'{o}n y Desarrollo (Chile), Ministerio de Ciencia, Tecnolog\'{i}a e Innovaci\'{o}n (Argentina), Minist\'{e}rio da Ci\^{e}ncia, Tecnologia, Inova\c{c}\~{o}es e Comunica\c{c}\~{o}es (Brazil), and Korea Astronomy and Space Science Institute (Republic of Korea). Processed using the Gemini IRAF package and DRAGONS (Data Reduction for Astronomy from Gemini Observatory North and South).

This work made use of data supplied by the UK Swift Science Data Centre at the University of Leicester.

The National Radio Astronomy Observatory is a facility of the National Science Foundation operated under cooperative agreement by Associated Universities, Inc.

This research is based on observations made with the NASA/ESA Hubble Space Telescope obtained from the Space Telescope Science Institute, which is operated by the Association of Universities for Research in Astronomy, Inc., under NASA contract NAS 5–26555. These observations are associated with program \#16923.

This work is partly based on observations made with the Gran Telescopio Canarias (GTC), installed at the Spanish Observatorio del Roque de los Muchachos of the Instituto de Astrofísica de Canarias, on the island of La Palma. Partly based on observations collected at the Calar Alto Astronomical Observatory, operated jointly by Instituto de Astrof\'isica de Andaluc\'ia (CSIC) and Junta de Andaluc\'ia.

Partly based on observations made with the Nordic Optical Telescope, under program 64-502, owned in collaboration by the University of Turku and Aarhus University, and operated jointly by Aarhus University, the University of Turku and the University of Oslo, representing Denmark, Finland and Norway, the University of Iceland and Stockholm University at the Observatorio del Roque de los Muchachos, La Palma, Spain, of the Instituto de Astrof\'isica de Canarias.

The LBT is an international collaboration among institutions in the United States, Italy and Germany. LBT Corporation partners are: The University of Arizona on behalf of the Arizona Board of Regents; Istituto Nazionale di Astrofisica, Italy; LBT Beteiligungsgesellschaft, Germany, representing the Max-Planck Society, The Leibniz Institute for Astrophysics Potsdam, and Heidelberg University; The Ohio State University, representing OSU, University of Notre Dame, University of Minnesota and University of Virginia.

\bmhead{Authors' Contributions}

J.C.R. is Principal Investigator of the MMT observations (shared P.I. with N.S. on MMIRS follow-up) and the {\em HST} program. J.C.R. reduced and analyzed the majority of the optical-NIR data and led the writing. B.P.G. identified the source as a possible merger, analyzed the high energy observations, provided modeling support and contributed to the text. A.J.L. analyzed observations, provided analysis and co-wrote the text. W.F. is Principal Investigator of the Gemini and VLA programs and provided input on analysis and text. M.N. performed the kilonova and the Ni-powered transient modeling, and contributed text. G.P.L. identified the source as a possible merger, devised joint kilonova and afterglow modeling method, and modeled the afterglow. D.B.M. is Principal Investigator of the NOT follow-up, and reduced and analyzed observations. A.E.N. reduced the Keck spectrum and performed stellar population modeling. S.R.O. analyzed the {\em Swift}/UVOT observations. N.R.T. provided input and rates analysis. A.d.U.P., D.A.K., J.F.A.F. and C.C.T. executed and reduced the CAHA and GTC observations. C.D.K reduced and analyzed the {\em HST} observations. C.J.M. calculated the GW observability. B.D.M., R.C., and M.E.R. provided input on modeling and analysis. A.R. and E.M. executed and reduced the LBT observation. G.S. executed and reduced the radio observation. J.J., D.J.S. and N.S. contributed MMT follow-up time and provided input on the scientific interpretation. L.I. and J.P.U.F. contributed to reduction of the NOT observations. A.E.N., P.K.B., C.D.K. and H.S. executed the Keck spectrum. E.B., R.C., B.E.C., M.D.P., T.L., K.P., A.R.C. are co-investigators of the programs used in this work and/or provided input on the scientific interpretation.

\noindent We declare no competing interests.

\noindent Correspondence and requests for materials should be addressed to Jillian Rastinejad.

\noindent  Reprints and permissions information is available at www.nature.com/reprints.

\noindent Supplementary Information is not available for this paper.

\noindent \textbf{Extended Data Figure 1. The host of GRB 211211A is a low mass, actively star forming galaxy in the local universe.} \textbf{(a)} The 2D NOT/ALFOSC spectra of the afterglow and host of GRB211211A. \textbf{(b)} Keck/DEIMOS 1D spectrum (blue) and $1\sigma$ uncertainty (dot-dashed blue line) compared with the arbitrarily-scaled NOT/ALFOSC afterglow spectrum (red), and \texttt{Prospector} model spectrum (grey). We highlight the strong emission lines in the observed host spectrum, none of which are detected in the 1D or 2D afterglow spectrum.  \textbf{(c)} The observed host photometry (blue circles) and $3\sigma$ uncertainties (blue lines), \texttt{Prospector} model photometry (black squares) and \texttt{Prospector} model spectrum (grey line). The \texttt{Prospector} derived SED matches the observed photometry, spectral continuum, and spectral line strengths well.

\noindent \textbf{Extended Data Fig. 2. Temporal evolution of the ultraviolet through near-IR spectral energy distribution (SED) of the counterparts to GRB\,211211A (solid lines) and GW170817 (dashed lines).} Circles or squares represent detections while triangles represent upper limits. For both GRB\,211211A and GW170817, the counterparts' SEDs at 4 -- 5.1~days post-burst (dark blue and purple) demonstrate a dramatic reddening compared to those at earlier epochs.

\noindent \textbf{Extended Data Figure 3. Corner plot showing posterior distributions for the basic kilonova model}. This model consists of three ejecta components and a fraction $\zeta$ of the blue (low-lanthanide) ejecta which is heated by shocks from the GRB jet. The final parameter is a white noise term for modeling systematics in the data. The labeled 1$\sigma$ error bars are statistical only; we estimate an additional systematic error of $\sim 50\%$ on these parameters (see Methods).

\noindent \textbf{Extended Data Figure 4. Light curve fit using the binary-based kilonova model \cite{Nicholl+21}.} The dashed lines show a model for AT\,2017gfo evaluated at the same redshift, $z=0.076$.

\noindent \textbf{Extended Data Figure 5. Corner plot showing posterior distributions for the binary-based kilonova model} The model consists of three ejecta components whose masses, velocities and opacities depend on the chirp mass and binary mass ratio ($q$) and the fraction of ejecta lost via disk ($\varepsilon$) and magnetic ($\alpha$) winds. A fraction $\zeta$ of the blue (low-lanthanide) ejecta is heated by shocks from the GRB jet over a timescale $t_{\rm shock}$. The final parameter is a white noise term for modeling systematics in the data. The labeled 1$\sigma$ error bars are statistical only; we estimate an additional systematic error of $\sim 50\%$ on these parameters (see Methods).

\noindent \textbf{Extended Data Figure 6. Light curve fit using a $^{56}$Ni-powered model.} This provides a poor fit, as the single radioactive component is unable to cool quickly enough to match the early UV and longer-term NIR emission. The best-fitting parameters require an unrealistic composition of 100\% $^{56}$Ni and an ejecta velocity pushing against the upper bound of the prior at $0.4c$.

\noindent \textbf{Extended Data Figure 7. The optical and near-IR light curves of GRB\,211211A have similar luminosities and decay rates compared to past kilonovae and kilonova candidates.} The rest-frame $i$-band (\textbf{a}) and $K$-band (\textbf{b}) light curves of GRB\,211211A (purple diamonds), GW170817/AT\,2017gfo (grey points; \cite{villar+17} and references therein) and previous short GRB kilonova upper limits (yellow triangles) and detections (yellow circles; \cite{lamb+19,troja+19,Fong+21,Rastinejad+21}). As there are no other rest-frame $K$-band kilonova light curves beyond AT\,2017gfo, we plot rest-frame $J$- and $H$-band SGRB kilonova observations for comparison (open circles and triangles; \cite{Fox+05,berger+13,Tanvir+13,lamb+19,troja+19,Fong+21,O'Connor+21,Rastinejad+21}). At $z=0.076$, the $K$-band counterpart to GRB\,211211A is of similar luminosity to AT\,2017gfo and fades on similar timescales.

\noindent \textbf{Extended Data Figure 8. The ejecta mass and velocities estimated for GRB\,211211A compared to those of past kilonovae and kilonova candidates.} Best-fit ejecta and velocity estimates (including 1$\sigma$ errors) of the red (\textbf{a}), purple (\textbf{b}) and blue (\textbf{c}) kilonova components of GRB\,211211A (purple boxes; Section~\ref{sec:kn_modeling}). We also plot ejecta mass and velocity estimates for two-component models of AT\,2017gfo  \cite[red boxes; compiled in ][ and references therein]{siegel19}, a three-component model of AT\,2017gfo \cite[red stars; ][]{Nicholl+21} and previous short GRB kilonovae \cite[labeled yellow boxes; ][]{barnes+16,lamb+19}. As two-component models of AT\,2017gfo do not distinguish between the ``purple'' and ``red'' components included in our analysis, we plot past two-component ``red'' estimates on both corresponding panels. We plot the dynamical ejecta estimates for GRB\,160821B on the red and blue panels and the disk mass on the purple panel. We plot the total estimate for GRB\,130603B on all panels. Our estimates for GRB\,211211A fall within the range of AT\,2017gfo and past kilonova candidates. As ejecta mass and velocity estimates are highly model-dependent, we note that the most robust comparison is between the three-component estimates for AT\,2017gfo (stars) and our results for GRB\,211211A.


\section*{Extended Data}
\setcounter{figure}{0}

\addtolength{\tabcolsep}{-2.6pt} 
\begin{table*}
\footnotesize
\begin{center}
\caption{Optical-Near-IR Observations of the Counterpart and Host Galaxy of GRB\,211211A
\label{tab:observations}}
\begin{tabular}{ccllccccc}
\toprule
$\delta t$	&
Filter &
Facility		&
Instrument &
$t_{\rm exp}$	&
Transient &
AG-subtracted$^{*}$ &
Host &
Ref.  \\
(days)		&
 &
 		&
 &
(s)		&
(AB mag) &
(AB mag) &
(AB mag) \\
\midrule
0.27 & g' & MITSuME & ... & 6600 & 20.34 $\pm$ 0.20 & 21.05 $\pm$ 1.07 & ... &  2 \\
0.27 & Rc & MITSuME & ... & 6600 & 20.26 $\pm$ 0.10 & 21.03 $\pm$ 1.15 & ... &  2 \\
0.27 & Ic & MITSuME & ... & 6600 & 20.37 $\pm$ 0.30 & 21.44 $\pm$ 1.95 & ... &  2 \\
0.43 & r & NEXT & ... & 2000 & 20.25 $\pm$ 0.07 & 20.51 $\pm$ 0.32 & ... & 3 \\
0.45 & z & NEXT & ... & 2400 & 19.88 $\pm$ 0.30 & 20.08 $\pm$ 0.39 & ... & 3 \\
0.46 & R & HCT & ... & 900 & 20.26 $\pm$ 0.13 & 20.49 $\pm$ 0.30 & ... &  4 \\
0.68 & i & CAHA & CAFOS & 2700 & 20.75 $\pm$ 0.08 & 20.92 $\pm$ 0.20 & ... & 1 \\
0.69 & g & NOT & ALFOSC & 240 & 21.00 $\pm$ 0.04 & 21.16 $\pm$ 0.18 & ... &  1 \\
0.69 & r & NOT & ALFOSC & 240 & 20.81 $\pm$ 0.05 & 20.97 $\pm$ 0.18 & ... &  1 \\
0.69 & i & NOT & ALFOSC & 240 & 20.89 $\pm$ 0.06 & 21.08 $\pm$ 0.21 & ... &  1 \\
0.70 & R & LCO & Sinistro & 1200 & 21.00 $\pm$ 0.09 & 21.18 $\pm$ 0.23 & ... & 5 \\
1.40 & r & GMG & ... & ... & $>$ 21.96 & $>$ 22.06 & ... & 6 \\
1.41 & R & DOT & 4Kx4K & ... & 21.83 $\pm$ 0.07 & 21.92 $\pm$ 0.12 & ... &  7 \\
1.43 & r' & GIT & ... & 1500 & $>$ 21.15 & $>$ 21.19 & ... & 4 \\
1.68 & i & CAHA & CAFOS & 2700 & 22.56 $\pm$ 0.13 & 22.70 $\pm$ 0.20 & ... & 1 \\
2.56 & Rc & Zeiss-1000 & ... & 3600 & $>$ 23.06 & $>$ 23.14 & ... & 8 \\
2.68 & i & CAHA & CAFOS & 2400 & 24.56 $\pm$ 0.34 & 24.64 $\pm$ 0.52 & ... & 1 \\
4.07 & K & Gemini & NIRI & 900 & 22.41 $\pm$ 0.14 & 22.45 $\pm$ 0.14 & ... & 1 \\
4.42 & R & DOT & 4Kx4K & ... & $>$ 23.87 & $>$ 23.93 & ... & 7 \\
4.70 & H & TNG & NICS & ... & $>$ 21.89 & $>$ 21.90 & ... & 9 \\
5.10 & K & Gemini & NIRI & 900 & 22.40 $\pm$ 0.17 & 22.42 $\pm$ 0.17 & ... & 1 \\
5.11 & i & Gemini & GMOS & 600 & 26.03 $\pm$ 0.30 & 26.51 $\pm$ 0.71 & ... & 1 \\
5.96 & J & MMT & MMIRS & 2400 & 24.17 $\pm$ 0.35 & 24.24 $\pm$ 0.33 & 19.00 $\pm$ 0.03 & 1 \\
6.08 & i & Gemini & GMOS & 1200 & $>$ 25.49 & $>$ 25.67 & ... & 1 \\
6.94 & K & MMT & MMIRS & 3600 & 23.43 $\pm$ 0.31 & 23.46 $\pm$ 0.28 & ... & 1 \\
7.98 & K & MMT & MMIRS & 2250 & 23.77 $\pm$ 0.30 & 23.81 $\pm$ 0.27 & 19.22 $\pm$ 0.07 & 1 \\
9.92 & K & MMT & MMIRS & 1170 & $>$ 22.11 & $>$ 22.12 & ... & 1 \\
17.65 & i & NOT & ALFOSC & 3000 & $>$ 24.67 & ... & ... & 1 \\
19.57 & i & CAHA & CAFOS & 4000 & $>$ 24.14 & ... & ... & 1 \\
46.94 & g & MMT & Binospec & 600 & $>$ 24.72 &  ... & 19.78 $\pm$ 0.06 & 1 \\
46.95 & r & MMT & Binospec & 600 & $>$ 24.48 &  ... & 19.42 $\pm$ 0.04 & 1 \\
46.97 & z & MMT & Binospec & 600 & $>$ 23.92 &  ...  & 19.18 $\pm$ 0.08 & 1 \\\
55.03 & i & Gemini & GMOS & 2640 & $>$ 26.77 &  ... & 19.16 $\pm$ 0.05 & 1 \\
65.95 & K$_s$ & GTC & EMIR & 3528 & $>$ 21.99 &  ...  & ... & 1 \\
88.82 & K$_s$ & LBT & LUCI & 7229 & $>$ 24.62 &  ...  & ... & 1 \\
97.85 & K & MMT & MMIRS & 3600 & $>$ 24.32 &  ...  & ... & 1 \\
122.18 & {\rm F140W} & HST & WFC3/IR & 2412 & $>$ 27.19 &  ...  & 18.95 $\pm$ 0.01 & 1 \\
123.54 & {\rm F606W} & HST & ACS/WFC & 2000 & $>$ 27.76 & ...  & 19.53 $\pm$ 0.01 & 1 \\
Arch. & u & SDSS &  ...  &  ...  &  ...  &  ...  & 20.86 $\pm$ 0.13 & 10 \\
Arch. & W1 & WISE &  ...  &  ...  &  ...  &  ...  & 19.76 $\pm$ 0.05 & 11 \\
\botrule
\multicolumn{9}{l}{ $^{*}$Magnitudes of the transient after subtracting the model afterglow flux as described in Methods.} \\ 
\multicolumn{9}{l}{ Magnitudes corrected for foreground Galactic extinction according to $A_V = 0.048$ mag \cite{SchlaflyFinkbeiner11}.} \\ 
\multicolumn{9}{l}{ All upper limits newly published in this work are given at the 3$\sigma$ level.} \\
\multicolumn{9}{l}{ References: (1) This work, (2) \cite{GCN31217}, (3) \cite{Xiao+22}, (4) \cite{GCN31227}, (5) \cite{GCN31214}, (6) \cite{GCN31232}, (7) \cite{GCN31299}, (8) \cite{GCN31234},} \\
\multicolumn{9}{l}{ (9) \cite{Mei+22}, (10) \cite{Alam+15}, (11) \cite{WISE_Wright+10}.} \\
\end{tabular}
\end{center}
\end{table*}
\addtolength{\tabcolsep}{2.0pt}

\begin{table*}
\begin{centering}
\caption{\textit{Swift}-UVOT Photometry of the Counterpart to GRB\,211211A
\label{tab:swift_obs}}
\begin{tabular}{ccccc}
\toprule
$\delta t$ & Band & $t_{\rm exp}$ & Transient & AG-subtracted$^{*}$ \\
(days) &  & (s) & (AB mag) & (AB mag) \\
\midrule
0.05 & $v$ & 199.8 & 19.17 $\pm$ 0.29 & 22.41 $\pm$ 2.50 \\
0.73 & $v$ & 189.8 & $>$ 19.21 & $>$ 19.24 \\
0.92 & $v$ & 186.6 & $>$ 19.20 & $>$ 19.22 \\
0.04 & $b$ & 199.8 & 19.65 $\pm$ 0.23 & $>$ 18.05 \\
0.20 & $b$ & 600.6 & 19.82 $\pm$ 0.19 & 20.46 $\pm$ 0.94 \\
0.78 & $b$ & 906.9 & 21.47 $\pm$ 0.48 & 21.66 $\pm$ 0.51 \\
1.12 & $b$ & 12553.7 & $>$ 20.63 & 22.08 $\pm$ 0.65 \\
0.04 & $u$ & 199.8 & 19.72 $\pm$ 0.15 & $>$ 18.13 \\
0.06 & $u$ & 80.0 & 19.45 $\pm$ 0.19 & 22.19 $\pm$ 2.50 \\
0.19 & $u$ & 906.8 & 19.76 $\pm$ 0.07 & 20.32 $\pm$ 0.75 \\
0.66 & $u$ & 184.5 & $>$ 20.36 & 22.09 $\pm$ 1.28 \\
0.86 & $u$ & 183.7 & $>$ 20.94 & $>$ 21.02 \\
1.19 & $u$ & 906.6 & $>$ 21.92 & $>$ 22.03 \\
0.06 & $uvw1$ & 199.8 & 19.47 $\pm$ 0.14 & 21.27 $\pm$ 2.50 \\
0.66 & $uvw1$ & 899.8 & 21.72 $\pm$ 0.24 & 21.99 $\pm$ 0.41 \\
0.85 & $uvw1$ & 899.8 & 21.91 $\pm$ 0.27 & 22.09 $\pm$ 0.34 \\
1.22 & $uvw1$ & 17535.0 & $>$ 21.98 & 23.07 $\pm$ 0.57  \\
0.05 & $uvm2$ & 199.7 & 19.59 $\pm$ 0.17 & 21.77 $\pm$ 2.50 \\
0.25 & $uvm2$ & 474.7 & 20.48 $\pm$ 0.17 & 21.06 $\pm$ 0.81 \\
1.26 & $uvm2$ & 826.4 & $>$ 22.18 & 23.89 $\pm$ 1.30 \\
0.05 & $uvw2$ & 199.8 & 19.61 $\pm$ 0.15 & 21.75 $\pm$ 2.50 \\
0.72 & $uvw2$ & 899.8 & 22.11 $\pm$ 0.26 & 22.38 $\pm$ 0.43\\
0.92 & $uvw2$ & 899.8 & $>$ 22.23 & 23.63 $\pm$ 0.94\\
0.00 & $white$ & 149.8 & 20.69 $\pm$ 0.24 & 24.11 $\pm$ 2.50 \\
0.79 & $white$ & 182.5 & 21.74 $\pm$ 0.35 & 21.95 $\pm$ 0.44 \\
1.06 & $white$ & 181.5 & 21.59 $\pm$ 0.36 & 21.69 $\pm$ 0.36 \\
1.78 & $white$ & 25045.5 & 23.41 $\pm$ 0.30 & 23.59 $\pm$ 0.37 \\
\hline
& & & Host \\
\hline
& $v$ &   & $>$ 19.78 & \\
& $b$ &  & 20.51 $\pm$ 0.14 & \\
& $u$ & &  21.21 $\pm$ 0.19 & \\
& $uvw1$ & &  21.66 $\pm$ 0.13 & \\
& $uvm2$ & & 21.86 $\pm$ 0.13 & \\ 
& $uvw2$ & &  21.87 $\pm$ 0.17 & \\
\bottomrule
\multicolumn{5}{l}{\tiny $^{*}$Magnitudes of the transient after subtracting the model afterglow flux as described in Methods.} \\ 
\multicolumn{5}{l}{\tiny Magnitudes corrected for foreground Galactic extinction according to $A_V = 0.048$ mag \cite{SchlaflyFinkbeiner11}.} \\ 
\end{tabular}
\end{centering}
\end{table*}

\begin{table*}
\begin{centering}
\caption{Afterglow Modeling Parameters
\label{tab:ag_params}}
\begin{tabular}{ccc}
\toprule
Parameter		&
Median &
Units \\
\midrule
log({$E_{\rm K, iso}$)} & $52.71^{+0.75}_{-0.78}$ & erg \\
$\Gamma_0$ & $73.11^{+51.70}_{-22.08}$ &  \\
$p$ & $2.014^{+0.007}_{-0.003}$ & \\
$\iota$ & $0.688^{+0.401}_{-0.344}$ & deg \\
{\rm log({$n$})} & $-0.265^{+1.289}_{-1.925}$ & cm$^{-3}$ \\
$\theta_c$ & $2.750^{+1.261}_{-1.432}$ & deg \\
{\rm log({$\varepsilon_e$})} & $-1.484^{+0.742}_{-0.807}$ & \\
{\rm log({$\mathit \epsilon_B$})} & $-3.819^{+1.535}_{-1.312}$ & \\
\botrule
\end{tabular}
\end{centering}
\end{table*}

\clearpage

\begin{figure*}
\centerline{
\includegraphics[width=1.0\textwidth]{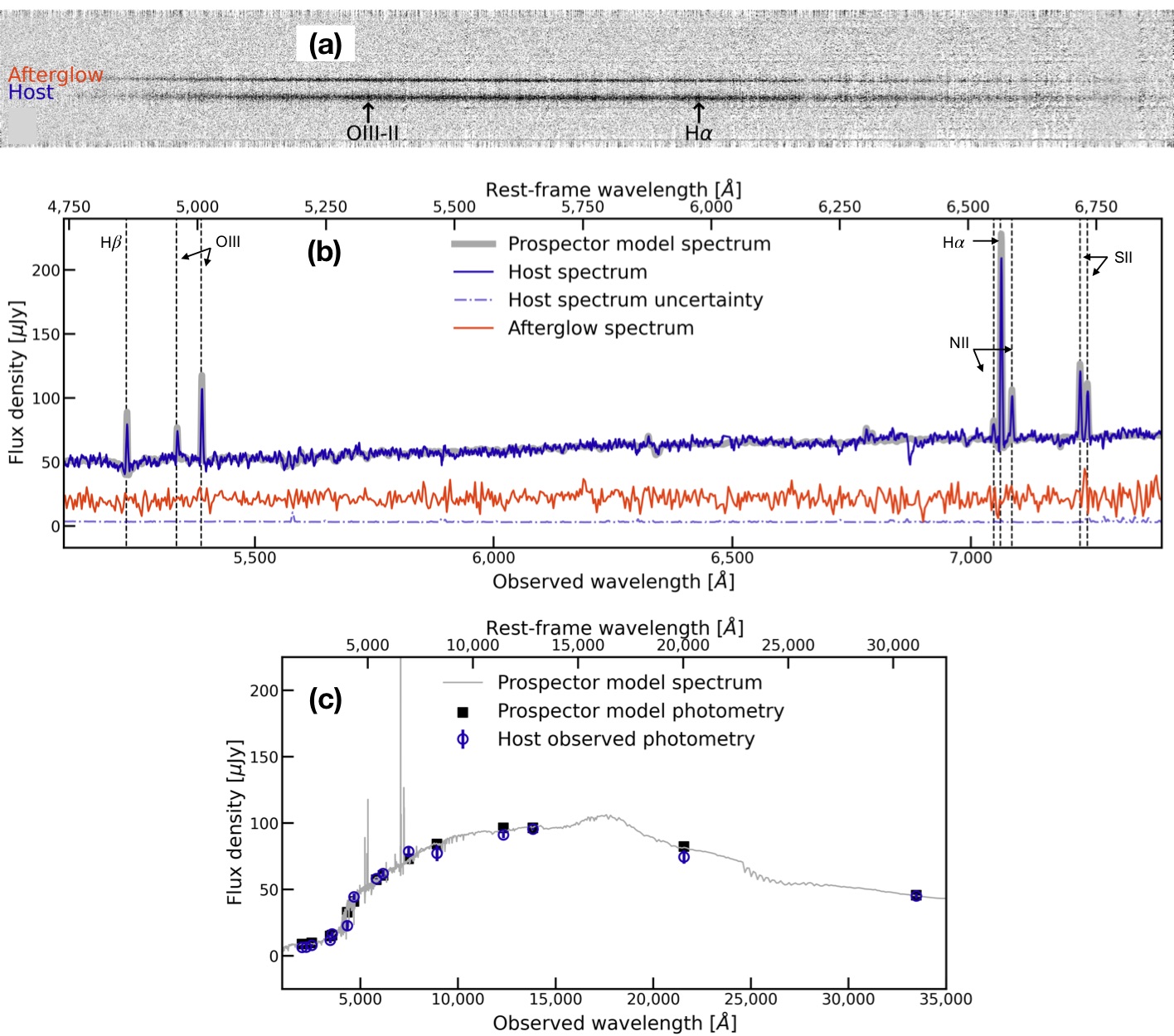}}
\caption{Extended Data}
\label{fig:prospectorSED}
\end{figure*}

\begin{figure*}
\centerline{
\includegraphics[width=\textwidth]{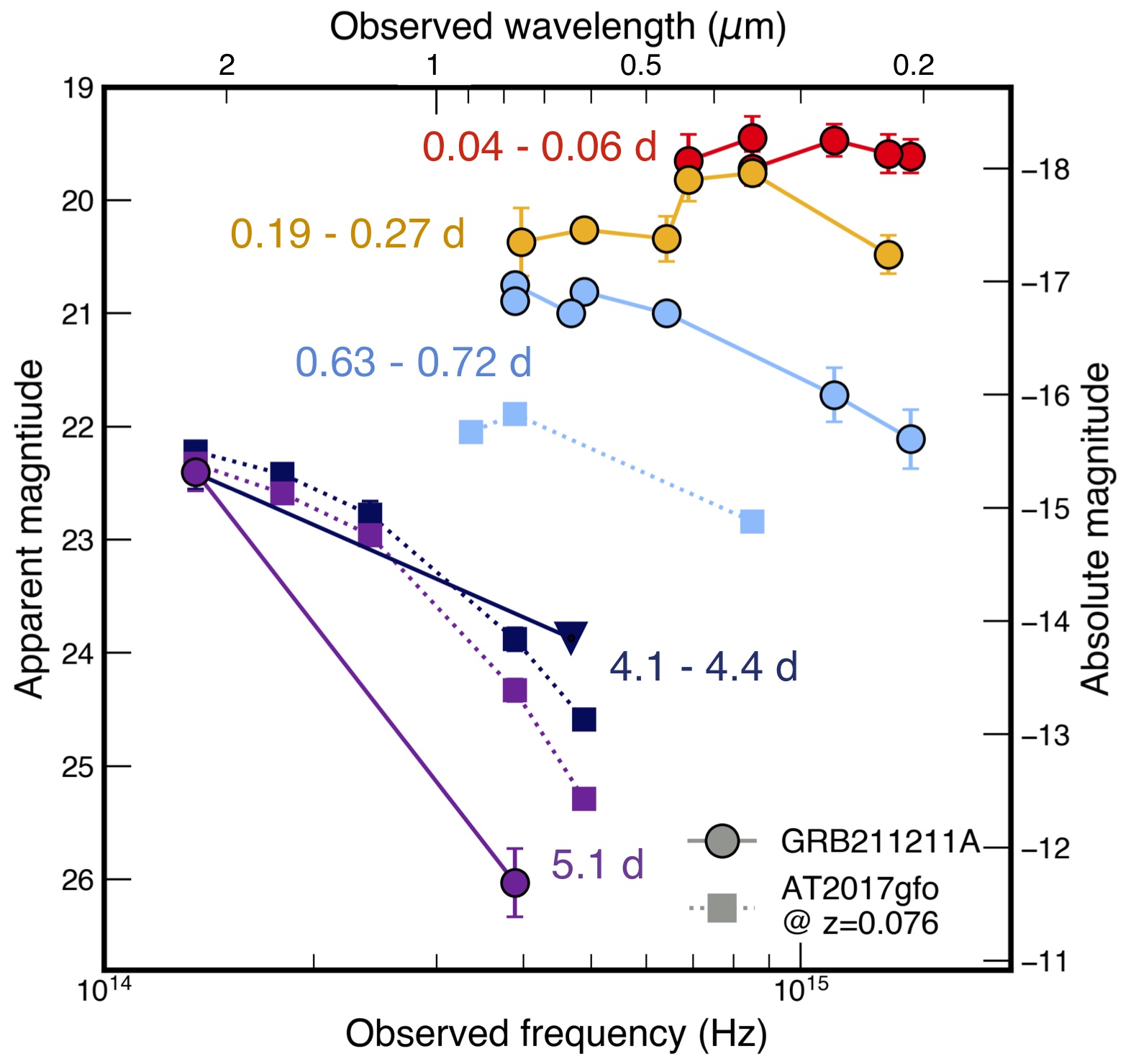}}
\caption{Extended Data}
\label{fig:uv_opt_nir_SED}
\end{figure*}

\begin{figure*}
\centerline{
\includegraphics[width=\textwidth]{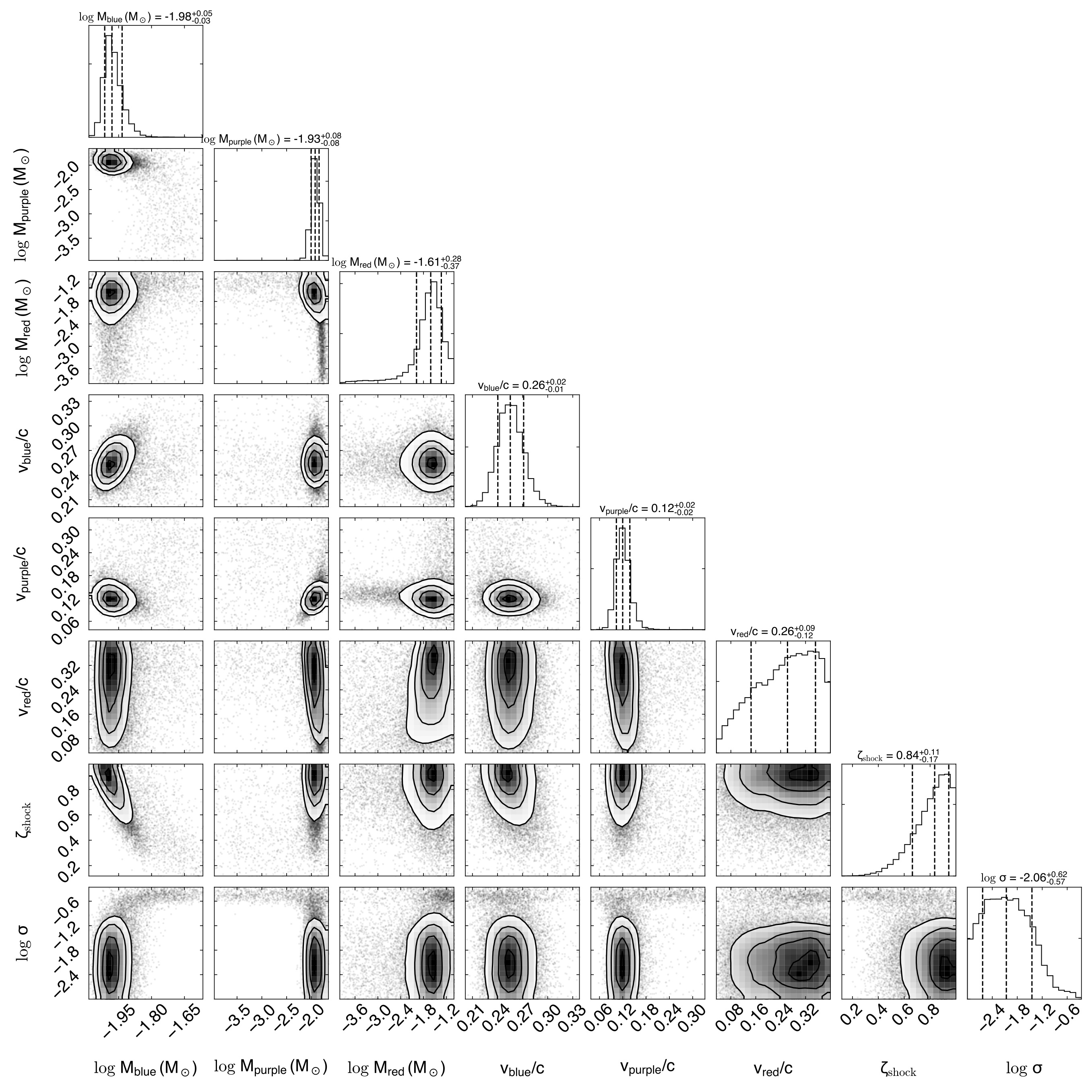}}
\caption{Extended Data}
\label{fig:KN_parameters}
\end{figure*}

\begin{figure*}
\centerline{
\includegraphics[width=.8\textwidth]{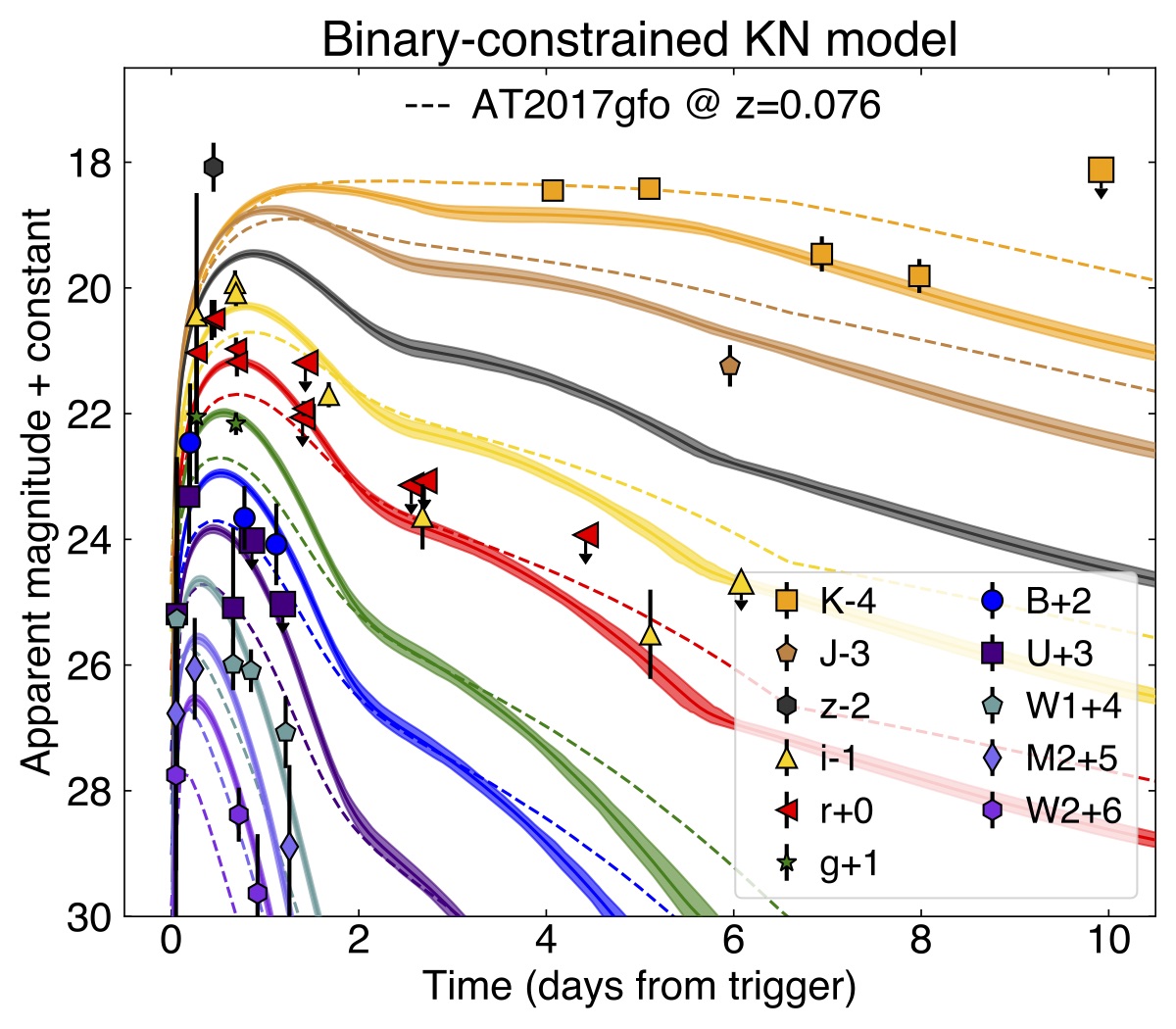}}
\caption{Extended Data.
}
\label{fig:binaryKN}
\end{figure*}

\begin{figure*}
\centerline{
\includegraphics[width=\textwidth]{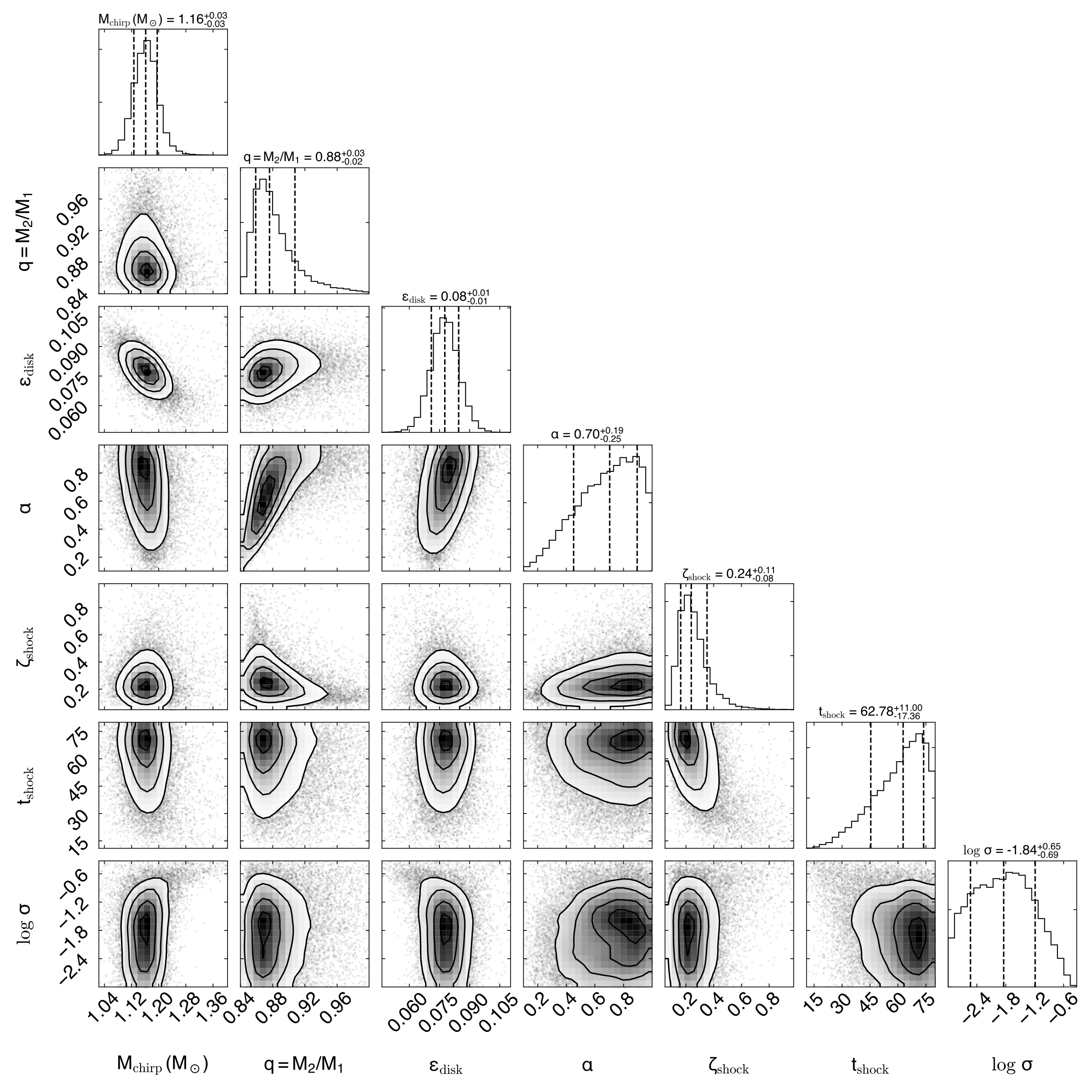}}
\caption{Extended Data.}
\label{fig:binaryKN_parameters}
\end{figure*}

\begin{figure*}
\centerline{
\includegraphics[width=.8\textwidth]{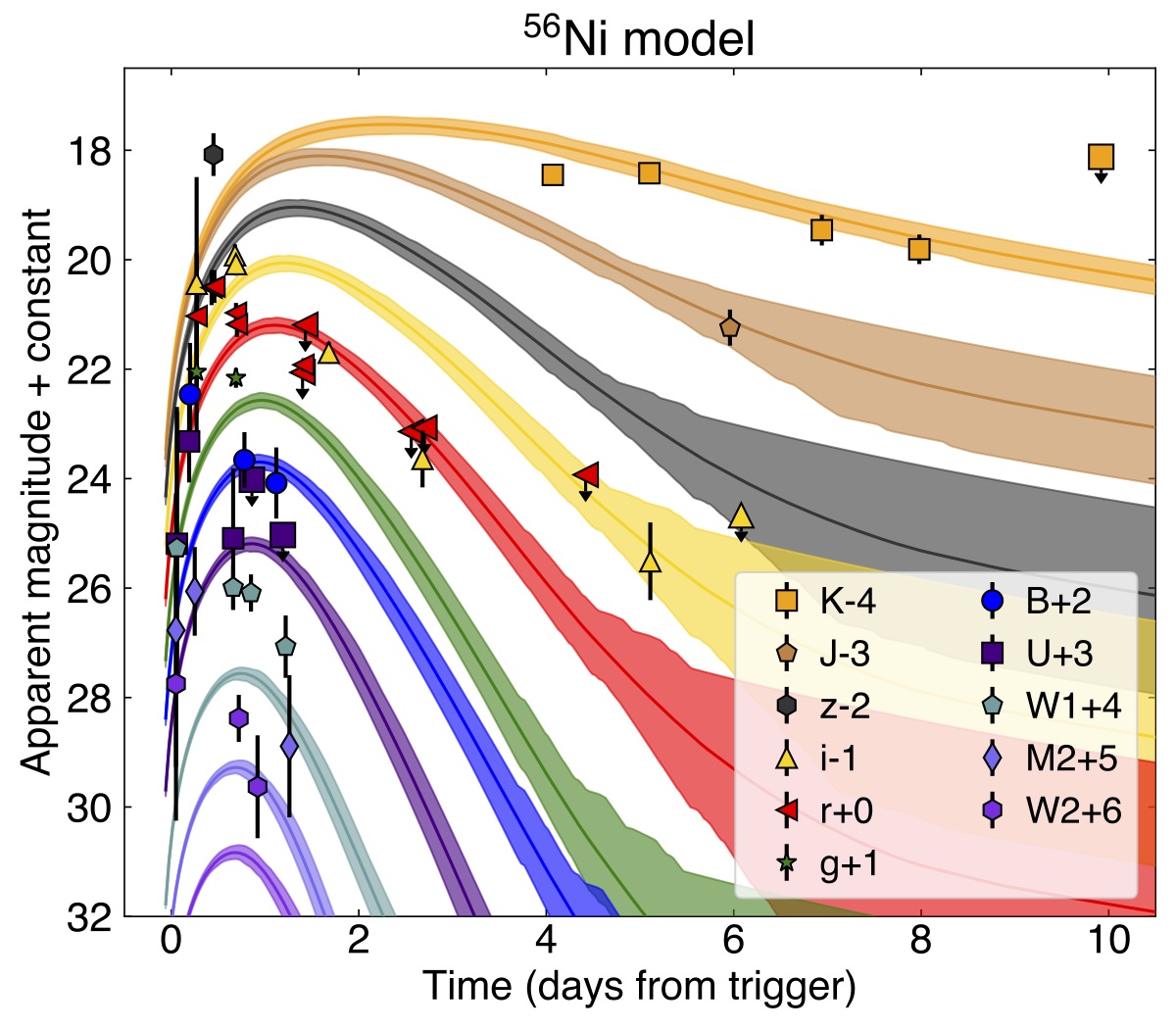}}
\caption{Extended Data.}
\label{fig:56ni_lc}
\end{figure*}

\begin{figure*}
\centerline{
\includegraphics[width=1.0\textwidth]{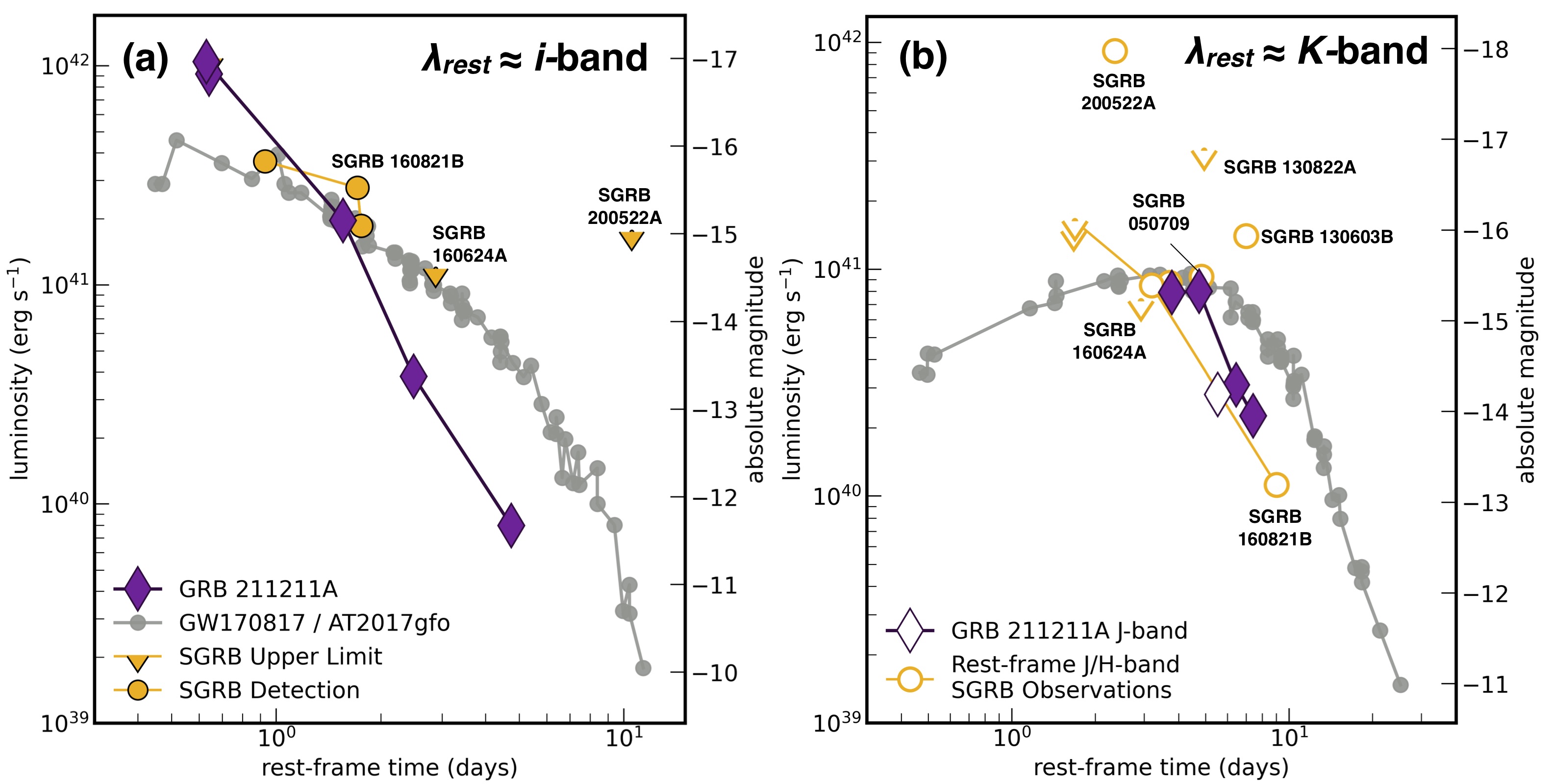}}
\caption{Extended Data.}
\label{fig:sgrb_lum_lc}
\end{figure*}

\begin{figure*}
\centerline{
\includegraphics[width=\textwidth]{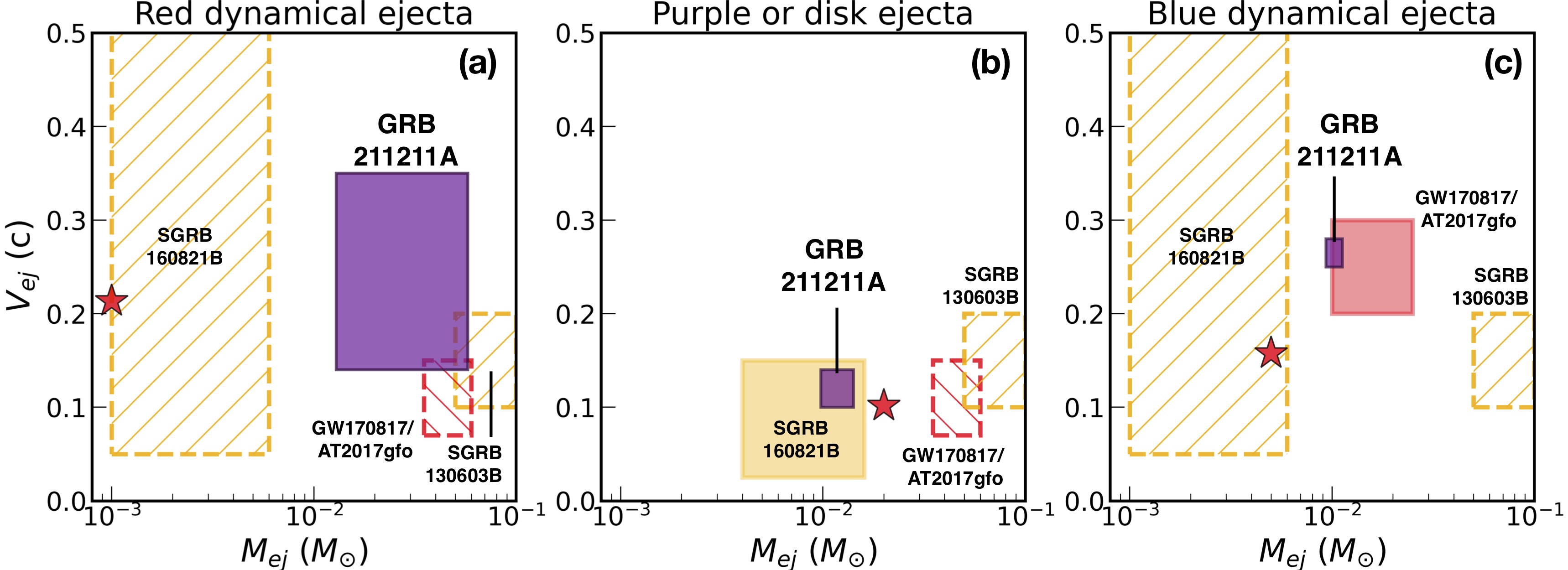}}
\caption{Extended Data.}
\label{fig:mej_vej}
\end{figure*}

\clearpage


\end{document}